%% file: quantum_tutorial_paper.tex
\documentclass[letterpaper, 10 pt,journal]{ieeetran}

\IEEEoverridecommandlockouts                              
\usepackage{amsmath} 
\usepackage{amssymb}  
\usepackage{braket}
\usepackage{graphicx}
\usepackage{mathtools}
\usepackage{tikz}
\usepackage{cite}
\usepackage{citesort}
\usepackage{blindtext}
\usepackage{tcolorbox}
\usetikzlibrary{decorations.pathreplacing}
\usetikzlibrary{shapes,arrows.meta}
\usetikzlibrary{calc}

\input{texdefs.tex}

\input{texdefs0.tex}
\input{texdefs1.tex}

\newcommand{\psiin}{\psi_{\rm in}}
\newcommand{\psif}{{\psi_{\rm out}}}
\newcommand{\Fuj}{F}
\newcommand{\Flb}{F_{\rm low-bnd}}
\newcommand{\delunc}{\del_{\rm unc}}
\newcommand{\Upert}{U_{\rm pert}}
\newcommand{\Unom}{U_{\rm nom}}
\renewcommand{\Fwc}{F_{\rm pert}}

\title{\LARGE \bf
Bringing Quantum Systems under Control:\\
{\large A Tutorial Invitation to Quantum Computing and Its Relation to Bilinear Control Systems}
}

\author{Julian Berberich$^1$, Robert L. Kosut$^2$, Thomas Schulte-Herbr\"uggen$^3$
\thanks{
   JB acknowledges funding by Deutsche Forschungsgemeinschaft (DFG, German Research Foundation) under Germany's Excellence Strategy - EXC 2075 - 390740016 and the support by the Stuttgart Center for Simulation Science (SimTech).
   RLK acknowledges support from The U.S. Army Research Office (ARO)
under contract No. W911NF-23-10307 and from the U.S. Department of
Energy (DOE) under STTR Contract No. DE-SC0020618.
TSH takes part in the {\em Munich Centre for Quantum Science and Technology} ({\sc mcqst}) and 
was supported i.a.\ by {\em Munich Quantum Valley} of the Bavarian State Government
with funds from Hightech Agenda Bayern Plus.}
\thanks{$^1$University of Stuttgart, Institute for Systems Theory and Automatic Control, 70569 Stuttgart, Germany {\tt\small julian.berberich@ist.uni-stuttgart.de}}%
\thanks{$^2$Systems \& Control Division, SC Solutions, Sunnyvale CA, USA, and Department of Chemistry, Princeton University, Princeton NJ, USA
{\tt\small kosut@scsolutions.com}
}%
\thanks{$^3$Technical University of Munich, School of Natural Sciences, 85748 Garching-Munich, Germany \&
Munich Quantum Valley (MQV), Schellingstr. 4, 80799 Munich, Germany {\tt\small tosh@tum.de}
}%
}

\begin{document}
\IEEEpubid{\begin{minipage}{\textwidth}\ \\[12pt] \\ \\
\copyright 2024 IEEE. This version has been accepted for publication in Proc. IEEE Conference on Decision and Control (CDC), 2024. Personal use of this material is permitted. Permission from IEEE must be obtained for all other uses, in any current or future media, including reprinting/republishing this material for advertising or promotional purposes, creating new collective works, for resale or redistribution to servers or lists, or reuse of any copyrighted component of this work in other works.
\end{minipage}}

\maketitle

\begin{abstract}
   Quantum computing
   comes with 
   the potential to push computational boundaries in various domains including, e.g., cryptography, simulation, optimization, and machine learning.
   Exploiting the principles of quantum mechanics, new algorithms can be developed with capabilities that are unprecedented by classical computers.
   However, the experimental realization of quantum devices is an active field of research with enormous open challenges, including robustness against noise and scalabilty.
   While systems and control theory plays a crucial role in tackling these challenges, the principles of quantum physics lead to a (perceived) high entry barrier for entering the field of quantum computing.
   
   This tutorial paper aims at lowering the barrier by introducing basic concepts required to understand and solve research problems in quantum systems. 
   First, we introduce 
   fundamentals of quantum algorithms, ranging from basic ingredients such as qubits and quantum logic gates to prominent examples and more advanced concepts, e.g., variational quantum algorithms.
   Next, we formalize some 
   engineering questions for building quantum devices 
in the real world, which requires the careful manipulation of microscopic quantities obeying quantum effects.
   To this end for $N$-level systems, we introduce basic concepts of (bilinear) quantum systems and control theory including controllability,  observability, and optimal control in a unified frame.
   Finally, we address the problem of noise in real-world quantum systems via robust quantum control, which relies on a set-membership uncertainty description frequently employed in control.

   A key goal of this tutorial paper is to demystify engineering aspects of quantum computing by emphasizing that its mathematical description mainly 
involves linear algebra (for quantum algorithms) and the handling of
bilinear control systems (for quantum systems and control theory) but does not require 
too much detailed knowledge 
of quantum physics.
\end{abstract}


\section{Introduction}
Quantum computing holds high potential for solving certain computational problems faster than with classical computers~\cite{nielsen2011quantum}.
Possible applications include integer factorization~\cite{shor1997polynomial}, which may be used to break current public-key encryption systems, unstructured search~\cite{grover1996fast}, and quantum simulation~\cite{lloyd1996universal}.
However, realizing these applications in experiments poses numerous challenges.
In particular, in the current noisy intermediate-scale quantum (NISQ) era~\cite{preskill2018quantum,bharti2022noisy}, quantum devices are fragile such that noise and uncertainty may cause significant perturbations.
While different approaches to address errors have been developed, in particular quantum error correction~\cite{gottesman2010introduction} and quantum error mitigation~\cite{cai2022quantum}, they do not (yet) allow to fully resolve errors on current quantum computers. 
\IEEEpubidadjcol

Systems and control theory plays a crucial role in realizing quantum devices.
In particular, experimental realizations require the development of specialized techniques for controlling quantum mechanical systems~\cite{dong2010quantum,altafini2012modeling,dong2022quantum,Koch22,DongPetersen2023}.
Also on the algorithmic level, various challenges in quantum computing are closely connected to control-theoretic concepts~\cite{berberich2023quantum}.
However, 
the principles of quantum physics lead to a (perceived) high entry barrier in particular also
for 
the engineering community to enter the field of quantum computing, and 
more widely, quantum technology.
This paper aims at lowering the barrier by providing a tutorial introduction to quantum systems.
Here we address the subject from three different, complementary perspectives, covering not only the algorithmic framework of quantum computers but also doability questions connected to quantum systems theory and robust control that arise in their experimental realization. --- Disclaimer: Naturally, there is a number of well-established important topics and results such a
didactic line-of-thought will have to skip. E.g., for infinite-dimensional systems, linear quantum control has successfully been used~\cite{JNP2008,dong2010quantum,DongPetersen2023}. In particular closed-loop feedback systems taking into account the backaction onto the quantum system via stochastic differential equations have been dwelled
upon extensively in~\cite{WisMil09}. Most notably, they were put to good use in the Nobel-prize winning setup of a photon box, where a chosen number of photons could be entered and stabilized~\cite{Haroche11,Haroche13}. For further reading see, e.g., the current European Quantum Control Roadmap~\cite{Koch22} and the references
therein.

For our goal, 
in Section~\ref{sec:quantum_algorithms} we first introduce the mathematical framework of quantum algorithms based on complex linear algebra.
We cover qubits, which are the main building blocks of quantum computers and are mathematically described by complex unit vectors, and we explain how they are accessed and manipulated via measurements and quantum gates, respectively.
Further, we discuss fundamental quantum algorithms, including variational quantum algorithms which are feedback interconnections of quantum and classical algorithms.

\newpage
Next, in Section~\ref{sec:tosh}, we sketch and exploit an emerging unified frame of quantum systems theory 
for the specific case of finite-dimensional bilinear quantum control systems
connecting to $N$-level systems and their gates in quantum information.
The frame is convenient and powerful for deciding engineering questions like controllability, observability, tomografiability etc. by common simple symmetry arguments.
We use established tools of linear algebra to absorb intricacies (of Lie theory) so as to remain within a familiar \/`matrix times vector\/' representation of linear maps now acting on operators (matrices) to exploit control on {\em orbits of quantum states}.
All relevant symmetries naturally show up in this frame---and breaking them by controls is key to get quantum systems under control. This guideline matches with numerical quantum optimal-control algorithms and helps to steer concrete experimental quantum setups in an optimized way, where modulated noise can be incorporated as additional control resource.
Bridges to powerful experimental settings are sketched.

Further, in Section~\ref{sec:robust_quantum_control}, we show how noise occurring in real-world applications can be addressed using ideas from classical robust control.
Specifically, we consider a set-membership uncertainty description for the underlying quantum system.
Expanding upon the classic Method of Averaging, the standard ``averaging transformation'' is modified to be suitable for quantum systems.
The resulting theory gives rise to an associated robust control optimization with two objectives, capturing a tradeoff between nominal performance and robustness.
Further, we present a fundamental bound on the robust performance of uncertain quantum systems.

Finally, Section~\ref{sec:conclusion} (Conclusions) wraps up the paper.

\input{part_JB}

\input{part_TSH}

\input{part_RK}

\section{Conclusion}\label{sec:conclusion}
This paper introduced algorithmic, systems- and control theoretical as well as experimental aspects of 
quantum engineering for
quantum computing from three complementary perspectives.
In Section~\ref{sec:quantum_algorithms}, we covered the framework of quantum algorithms, which are formulated in the language of linear algebra as a combination of unit vectors, unitary matrices, and quadratic forms.
Many 
experimental realizations towards 
a quantum computer or other high-end quantum technolgies (in finite dimensions)
require to control (bilinear) systems with high accuracy.
With this motivation, Section~\ref{sec:tosh} presented a unified systems and control framework for such quantum systems, using their symmetries to simplify rigorous treatment of decision problems 
such as controllability, simulability, and observability.
Finally, current quantum devices are severely affected by noise, which requires the development of robust techniques.
To this end, Section~\ref{sec:robust_quantum_control} introduced a robust quantum-control framework based on a set-membership uncertainty description and an optimal control objective trading off performance and robustness.

The main goal of this tutorial is to introduce basic concepts 
to understand and solve 
engineering 
problems in 
quantum systems.
Throughout the paper, we have encountered various open challenges in quantum computing which are closely connected to control-theoretic concepts.
We hope that addressing these challenges based on principles outlined in this tutorial will contribute towards \emph{bringing quantum systems under control}.

\bibliographystyle{IEEEtran}   
\bibliography{Literature, control21OLD}  

\end{document}

%% file: texdefs.tex
\newcommand{\mrm}{\rm}

\newcommand{\nb}{ {n^{}_B} }

\newcommand{\ns}{{n^{}_S}}

\newcommand{\btab}{\begin{tabular}}
\newcommand{\etab}{\end{tabular}}

\newcommand{\rank}{{\bf rank}}

\newcommand{\avg}{{\bf E}}

\newcommand{\tr}{{\mrm tr}~}

\renewcommand{\th}{\theta}
\newcommand{\del}{\delta}
\newcommand{\sig}{\sigma}

\newcommand{\norm}[1]{ \left\| #1 \right\| }

\newcommand{\eg}{\emph{e.g.}}
\newcommand{\ie}{\emph{i.e.}}
\newcommand{\etc}{\emph{etc.}}

\newcommand{\bquem}{\begin{quote}\begin{em}}
\newcommand{\equem}{\end{em}\end{quote}}

\newcommand{\blist}{\begin{description}}
\newcommand{\elist}{\end{description}}

\newcommand{\bquote}{\begin{quote}}
\newcommand{\equote}{\end{quote}}

\newcommand{\ben}{\begin{enumerate}}
\newcommand{\een}{\end{enumerate}}

\newcommand{\bit}{\begin{itemize}}
\newcommand{\eit}{\end{itemize}}

\newcommand{\bea}{\begin{array}}
\newcommand{\eea}{\end{array}}

\newcommand{\bds}{\begin{displaystyle}}
\newcommand{\eds}{\end{displaystyle}}

\newcommand{\ds}{\displaystyle}

\makeatletter
\def\beq{\@ifnextchar 
[{\@tempswatrue\@beq}{\@tempswafalse\@beq[]}}
\def\@beq[#1]{\begin{equation}\edef\@tmparg{#1}\ifx\@tmparg\@e
mpty \else
	\label{#1}\fi}
\newcommand{\eeq}{\end{equation}}
\makeatother 

\newcommand{\beqaa}{\begin{eqnarray*}}
\newcommand{\eeqaa}{\end{eqnarray*}}

\newcommand{\beqa}{\begin{eqnarray}}
\newcommand{\eeqa}{\end{eqnarray}}

\newcommand{\bc}{\begin{center}}
\newcommand{\ec}{\end{center}}

\newcommand{\bfig}{\begin{figure}}
\newcommand{\efig}{\end{figure}}

\renewcommand{\ns}{{n}}

\renewcommand{\rank}{\mbox{\rm rank}}


%% file: texdefs0.tex
\usepackage[space]{grffile}

\newcommand{\ffrefs}{\cite{GreenETAL:2013,Soare2014,Kaby:2014,Paz-Silva2014,QCTRL:2021,HaasHamEngr:2019,Chalermpusitarak2021,Cerfontaine2021}}

\newcommand{\toprefs}{\cite{RabitzHR:04,Ho.PRA.79.013422.2009,BeltraniETAL:2011,MooreR:2012,HockerETAL:2014,KosutArenzRabitz:2019}}

\newcommand{\Jrbst}{J_{\rm rbst}}

\renewcommand{\ns}{{n_S}}
\renewcommand{\nb}{{n_B}}

\newcommand{\Hcalunc}{{\cal H}_{\rm unc}}

\renewcommand{\th}{\theta}
\newcommand{\Hbf}{{\cal H}}

\newcommand{\Gavg}{G_{\rm avg}}

\newcommand{\Fnom}{F_{\rm nom}}
\newcommand{\Fwc}{F_{\rm wc}}

\newcommand{\normsm}[1]{\|#1\|}

\newcommand{\beasep}[1]{\renewcommand{\arraystretch}{#1}\bea}

\newcommand{\red}[1]{#1}

\newcommand{\blue}[1]{#1}

%% file: texdefs1.tex
\makeatletter
\newcommand*\rel@kern[1]{\kern#1\dimexpr\macc@kerna}
\newcommand*\widebar[1]{%
  \begingroup
  \def\mathaccent##1##2{%
    \rel@kern{0.8}%
    \overline{\rel@kern{-0.8}\macc@nucleus\rel@kern{0.2}}%
    \rel@kern{-0.2}%
  }%
  \macc@depth\@ne
  \let\math@bgroup\@empty \let\math@egroup\macc@set@skewchar
  \mathsurround\z@ \frozen@everymath{\mathgroup\macc@group\relax}%
  \macc@set@skewchar\relax
  \let\mathaccentV\macc@nested@a
  \macc@nested@a\relax111{#1}%
  \endgroup
}
\makeatother

\newcommand{\Hb}{\bar{H}}
\newcommand{\Ht}{\tilde{H}}

\renewcommand{\ns}{{N_S}}
\renewcommand{\nb}{{N_B}}

\renewcommand{\avg}[1]{\langle #1 \rangle}

\renewcommand{\rank}{\mbox{rank }}

\newtheorem{thm}{Theorem}

\renewcommand{\th}{\theta}

\newcommand{\Hcal}{{\cal H}}

%% file: part_JB.tex

\section{Fundamentals of quantum algorithms}\label{sec:quantum_algorithms}

The development of powerful quantum algorithms has been a key driver behind the recent surge of research in quantum technologies.
In this section, we provide a tutorial introduction to the mathematical framework of quantum algorithms based on complex linear algebra. 
In particular, we cover basic concepts such as qubits (Section~\ref{subsec:qubits}), measurement (Section~\ref{subsec:measurement}), and quantum gates (Section~\ref{subsec:gates}).
We then combine these concepts into quantum algorithms in Section~\ref{subsec:algorithms}, introducing both the main mathematical concept as well as several important examples.
Next, in Section~\ref{subsec:vqas}, we introduce variational quantum algorithms (VQAs) which are feedback interconnections of classical and quantum algorithms.
Finally, we discuss density matrices which provide an alternative and often useful mathematical description of quantum states (Section~\ref{subsec:density_matrices}).

This section focuses on the fundamental basics of quantum algorithms and we refer to~\cite{berberich2023quantum} for a more comprehensive introduction from the control perspective as well as to~\cite{nielsen2011quantum} for an excellent standard primer 
on quantum computing.

\subsection{Qubits}\label{subsec:qubits}

Quantum bits (\emph{qubits} in short) are the basic unit of computation on a quantum computer.
Mathematically, a qubit is a complex unit vector $\ket{\psi}\in\mathbb{C}^2_{=1}$, where we use the standard \emph{Dirac notation} $\ket{\psi}$ for quantum states.
It is insightful to express a qubit in the \emph{computational basis} as 
\begin{align}
    \ket{\psi}=\alpha\ket0+\beta\ket1
\end{align}
for the basis states $\ket0=\begin{bmatrix}
    1\\0
\end{bmatrix}$, $\ket1=\begin{bmatrix}
    0\\1
\end{bmatrix}$.
Here, $\alpha$ and $\beta$ are commonly referred to as the probability amplitudes.
Whereas a classical bit takes one of two values $0$ or $1$, a qubit lies in \emph{superposition} of the basis states $\ket0$ and $\ket1$.
A measurement of the qubit in the computational basis collapses this superposition onto one of two possible outcomes:
either $0$ or $1$ with probability $|\alpha|^2$ and $|\beta|^2$, respectively (see Section~\ref{subsec:measurement} for details).

A quantum state composed of $n$ qubits is a $2^n$-dimensional complex unit vector, i.e., $\ket{\psi}\in\mathbb{C}^{2^n}_{=1}$.
The computational basis for this space is formed by $n$-fold tensor products (i.e., Kronecker products) of the basis states $\ket0$ and $\ket1$ with themselves, i.e., 
\begin{align}\label{eq:computational_basis_states}
    \ket{0\dots00},\>\>\ket{0\dots01},\>\>\text{etc.},
\end{align}
where we use the common notation $\ket{ab}=\ket{a}\otimes\ket{b}$ for bit strings $a$ and $b$.
A simple example of a $2$-qubit quantum state is the tensor product of two single-qubit states $\ket{\psi_1}$ and $\ket{\psi_2}$, i.e., $\ket{\psi}=\ket{\psi_1}\otimes\ket{\psi_2}$.
States which can be written as the tensor product of two quantum states are called \emph{separable}.
It is an important fact that not all multi-qubit states are separable, e.g., the \emph{Bell state} 
\begin{align}\label{eq:bell_state_def}
\ket{\Phi^+}=\tfrac{1}{\sqrt{2}}(\ket{00}+\ket{11})
\end{align}
cannot be decomposed into a tensor product of two single-quit states.
States which are not separable are called \emph{entangled}.
Entanglement corresponds to a strong form of correlation 
which is unique to quantum systems and a key ingredient for computational speedups of quantum algorithms~\cite{jozsa2003role}.

\subsection{Projective Measurements}\label{subsec:measurement}

Contrary to classical physics, it is in general not possible to precisely determine the state of a quantum system.
In the following, we introduce \emph{projective measurements}, which allow to retrieve information about quantum states according to certain rules (see~\cite{nielsen2011quantum} for details and alternative forms of measurements).

Projective measurements are taken w.r.t.\ an \emph{observable}, i.e., a Hermitian matrix $M=M^\dagger\in\mathbb{C}^{2^n\times 2^n}$, where we use $M^\dagger$ to denote the Hermitian transpose of a matrix $M$.
Let us write $M$ in its spectral decomposition
\begin{align}
    M=\sum_i\lambda_iP_i
\end{align}
with the different eigenvalues $\lambda_i$ and the orthogonal projectors $P_i$ onto the corresponding eigenspaces.
When performing a projective measurement of the state $\ket{\psi}$ w.r.t.\ $M$, the outcome is always one of the $\lambda_i$'s.
A particular $\lambda_i$ is returned with probability 
\begin{align}\label{eq:measurement_probability}
    \braket{\psi|P_i|\psi}\coloneqq\psi^\dagger P_i\psi.
\end{align}
Moreover, immediately after the measurement, the quantum state $\ket{\psi}$ \emph{collapses}, i.e., it changes its value to the 
orthogonal projection on the corresponding eigenspace
%
\begin{align}
    \frac{P_i\ket\psi}{\braket{\psi|P_i|\psi}}.
\end{align}
The fact that quantum states possibly change their values due to a measurement poses unique challenges to the design and implementation of quantum algorithms.

Let us illustrate the concept of projective measurements with a simple example:
Suppose we measure a single qubit $\ket{\psi}=\alpha\ket0+\beta\ket1$ w.r.t.\ the Pauli matrix $Z=\begin{bmatrix}
    1&0\\0&-1
\end{bmatrix}$, an operation that is commonly referred to as measurement in the computational basis.
According to the above exposition, the measurement can only return one of the two eigenvalues $\lambda_{1/2}=\pm1$ of $Z$.
The probability for measuring $\lambda_1=+1$ or $\lambda_2=-1$ is 
\begin{align}
    \braket{\psi|P_1|\psi}&=|\alpha|^2\quad
    \text{and}\quad\braket{\psi|P_2|\psi}=|\beta|^2,
\end{align}
respectively,
where $P_1=\begin{bmatrix}
    1&0\\0&0
\end{bmatrix}$ and $P_2=\begin{bmatrix}
    0&0\\0&1
\end{bmatrix}$.
Moreover, if the measurement returns $+1$, then, immediately afterwards, the quantum state is equal to the corresponding eigenvector $\ket0$, and likewise for the measurement outcome $-1$ with the corresponding eigenvector $\ket1$.

In many practical scenarios, one is not interested in a particular measurement outcome but, instead, it is desirable to perform repeated measurements of the state $\ket\psi$ in order to determine the value of the quadratic form
\begin{align}\label{eq:meas_quad_form}
    \braket{\psi|M|\psi}=\psi^\dagger M\psi.
\end{align}
Since measurements affect the state in a possibly undesirable fashion, this requires the availability of independent copies of $\ket\psi$ which need to be generated separately.
Assuming that such copies are available, a statistical estimate of~\eqref{eq:meas_quad_form} can be determined based on projective measurements.
To be precise, suppose we have $T$ independent copies of $\ket\psi$ and measure each state w.r.t.\ $M$, obtaining each outcome $\lambda_i$ for $T_i$ of these measurements.
Then, we can form the approximation 
\begin{align}\label{eq:measurement_approximation}
    \braket{\psi|M|\psi}\approx\frac{\sum_i\lambda_iT_i}{T}.
\end{align}
Projective measurements can be interpreted as sampling from a Bernoulli distribution.
This allows to derive error bounds for the approximation in~\eqref{eq:measurement_approximation}, which vanish when $T\to\infty$, see~\cite[Section 3.2.4]{schuld2021machine} for details.
To summarize, projective measurements can be interpreted as statistical estimates of the quadratic form~\eqref{eq:meas_quad_form}.

\subsection{Quantum Gates}\label{subsec:gates}

\begin{figure}
    \begin{center}
        \includegraphics[width=0.75\columnwidth]{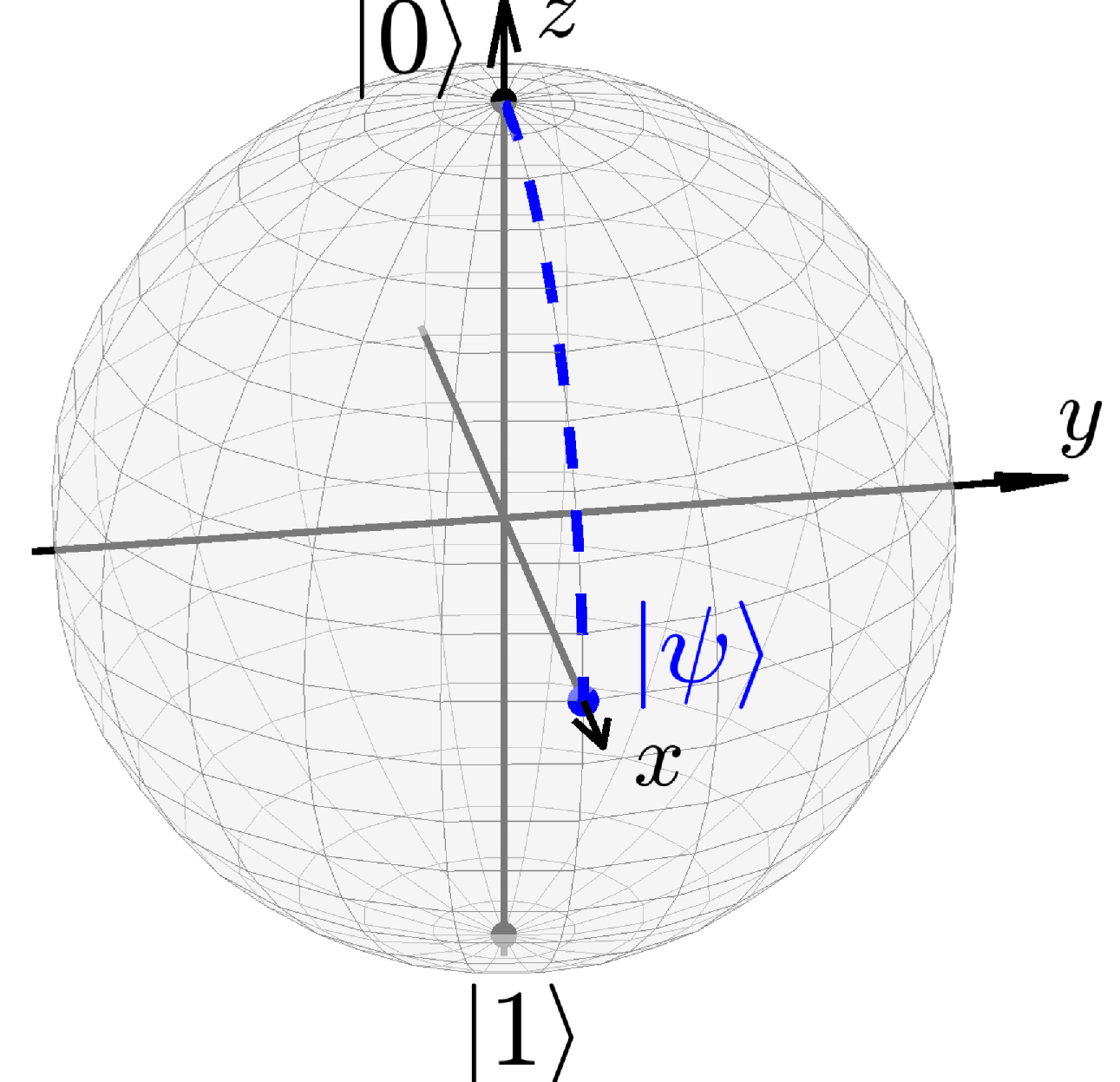}
    \end{center}
    \caption{Bloch-sphere representation of a single-qubit rotation $R_\mathrm{y}(\frac{\pi}{2})$ applied to the qubit in state $\ket0$.}
    \label{fig:bloch_sphere}
    \end{figure}

On a quantum computer, qubits are manipulated using \emph{quantum gates}, which are the quantum analog to classical logical gates (AND, NOT, OR, etc.).
Mathematically, quantum gates are unitary matrices $U\in\mathbb{C}^{2^n\times 2^n}$, i.e., they satisfy $U^\dagger U=I$.
A quantum gate $U$ acts on a qubit via multiplication, i.e., applying $U$ to $\ket{\psi}$ produces $U\ket{\psi}$.
Notable examples of quantum gates include the Pauli gates 
\begin{align}
    X=\begin{bmatrix}
        0&1\\1&0
    \end{bmatrix},\>\>
    Y=\begin{bmatrix}
        0&-i\\i&0
    \end{bmatrix},\>\>
    Z=\begin{bmatrix}
        1&0\\0&-1
    \end{bmatrix}.
\end{align}
The Pauli-$X$ gate is frequently referred to as NOT since $X\ket0=\ket1$ and $X\ket1=\ket0$.
Intuitively, quantum gates can be thought of as rotations.
To show this, we resort to the \emph{Bloch sphere}, which is a useful graphical illustration of qubits.
Since multiplication of a quantum state by a \emph{global phase} $e^{-i\varphi}$ does not affect the outcome of a measurement (compare~\eqref{eq:measurement_probability}), quantum states which only differ by a global phase are considered equivalent.
Using additionally that $\lVert\ket{\psi}\rVert=1$, a qubit $\ket{\psi}$ has two free parameters, which can be mapped onto a sphere in $\mathbb{R}^3$, compare Figure~\ref{fig:bloch_sphere}.
The north pole and south pole of this sphere is $\ket0$ and $\ket1$, respectively, whereas intermediate values are in superposition of these two basis states.

Figure~\ref{fig:bloch_sphere} illustrates a simple quantum gate:
the rotation around the $y$-axis by angle $\frac{\pi}{2}$, which corresponds to the unitary matrix $R_{\mathrm{y}}(\frac{\pi}{2})$ for the \emph{rotation gate}
\begin{align}
    R_{\mathrm{y}}(\theta)=e^{-i\frac{\theta}{2}Y}.
\end{align}

\begin{figure}
    \begin{center} 
\newcommand\linelength{0.5}
\newcommand\verticaldiff{0.5}
\begin{tikzpicture}[scale = 1]

\node (lu) at (0,0) {};
\node (rl) at ($(lu)+(1.2,-1)$) {};
\draw ($(lu)$) rectangle ($(rl)$);
\node at ($(lu)+(0.6,-0.5)$) {\Large $U$};

\draw ($(lu)+(-\linelength,-\verticaldiff)$) -- ($(lu)+(0,-\verticaldiff)$);
\node at ($(lu)+(-\linelength-0.65,-\verticaldiff)$) {\Large $\ket{\psi}$};
\draw ($(rl)+(\linelength,\verticaldiff)$) -- ($(rl)+(0,\verticaldiff)$);
\node at ($(rl)+(\linelength+0.75,\verticaldiff)$) {\Large $U\ket{\psi}$};
\end{tikzpicture}%
\end{center}
    \caption{Circuit representation of a quantum gate $U$ acting on the input state $\ket{\psi}$ and leading to the output state 
    $U\ket{\psi}$.}
    \label{fig:quantum_gate}
\end{figure}
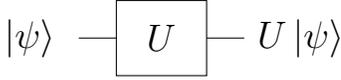

To design a non-trivial quantum algorithm, multi-qubit gates are required.
A popular example is the \emph{controlled-NOT} gate (CNOT), which is a $2$-qubit gate defined via the unitary matrix 
\begin{align}
    \mathrm{CNOT}=\left[\begin{smallmatrix}
        1&0&0&0\\
        0&1&0&0\\
        0&0&0&1\\
        0&0&1&0
    \end{smallmatrix}\right].
\end{align}
When applied to a quantum state of the form $\ket{\psi_1}\otimes\ket{\psi_2}$, the CNOT yields 
\begin{align}
    &\ket{\psi_1}\otimes \ket{\psi_2}\quad\quad\text{if\>$\ket{\psi_1}=\ket0$}\\
    \text{and}\quad&\ket{\psi_1}\otimes X\ket{\psi_2}\quad\text{if\>$\ket{\psi_1}=\ket1$},
\end{align}
i.e., a Pauli-$X$ gate (the NOT gate) is applied to the second qubit if the first qubit is in state $\ket1$.
In quantum algorithms, the CNOT gate is typically used to generate entanglement.
For example, when initializing two qubits in the computational basis state $\ket{00}$, applying the \emph{Hadamard gate} $H=\frac{1}{\sqrt{2}}\begin{bmatrix}1&1\\1&-1\end{bmatrix}$ to the first qubit and then applying a CNOT yields the prime example of an entangled state, the Bell state~\eqref{eq:bell_state_def}, i.e.,
\begin{align}\label{eq:Bell_state_circuit}
    \ket{\Phi^+}=\mathrm{CNOT}(H\otimes I_2)\ket{00},
\end{align}
where $I_2$ is a two-dimensional identity matrix.
Quantum gates are commonly illustrated in their circuit representation, compare Figure~\ref{fig:quantum_gate}.

\subsection{Quantum Algorithms}\label{subsec:algorithms}

\begin{figure}
    \begin{center} 
\begin{tikzpicture}
\draw (-2,0) -- (2,0);
\draw (0,1) -- (2,1);    
\draw (1,1) -- (1,-0.2);
\filldraw (1,1) circle (0.1cm);
\draw (1,0) circle (0.2cm);    
\node (lu) at (-1.2,1.5) {};
\node (rl) at ($(lu)+(1.2,-1)$) {};
\draw ($(lu)$) rectangle ($(rl)$);
\node at ($(lu)+(0.6,-0.5)$) {\Large $H$};
\node at (-2.55,1) {\Large $\ket{0}$};
\node at (-2.55,0) {\Large $\ket{0}$};
\draw (-2,1) -- (-1.2,1);
\node at (2.26,0) {\includegraphics[width=0.25in]{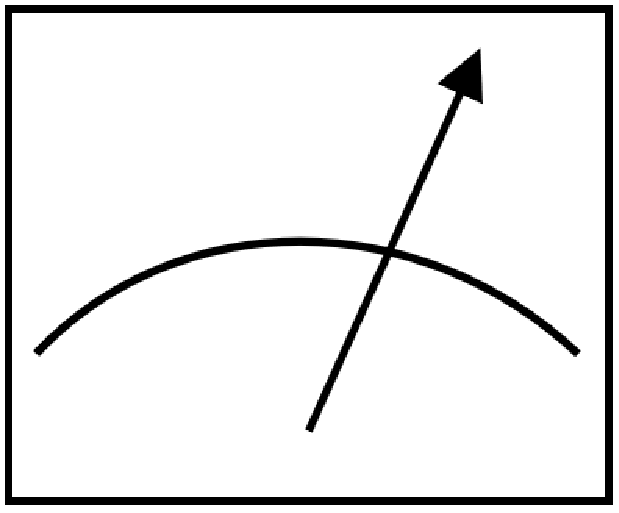}};
\node at (2.26,1) {\includegraphics[width=0.25in]{measurement}};
\end{tikzpicture}%
\end{center}
\caption{Quantum circuit implementing the algorithm in~\eqref{eq:Bell_state_circuit} generating a Bell state $\ket{\Phi^+}$, followed by a measurement of each individual qubit w.r.t.\ the observable $Z$.
}\label{fig:quantum_gate_bell_state}
\end{figure}

A quantum algorithm typically is a combination of qubits, quantum gates, and measurements.
Quantum algorithms are commonly displayed in their circuit representation and are, therefore, also referred to as quantum circuits.
Figure~\ref{fig:quantum_gate_bell_state} illustrates the circuit implementing the operation in~\eqref{eq:Bell_state_circuit}, i.e., converting the initial state $\ket{00}$ to the Bell state $\ket{\Phi^+}$, which is then measured.
Measurements are generally depicted via the \emph{meter} symbol on the right end of the circuit.
If no observable is mentioned explicitly, then the measurement is understood as measurement in the computational basis, i.e., measuring each individual qubit w.r.t.\ $Z$.

By convention, quantum circuits are read from left to right.
It is simple to show that a parallel and series interconnection of unitary matrices is, again, a unitary matrix.
Hence, we obtain the following statement.\footnote{Throughout this tutorial paper, colored boxes contain key statements and takeaway messages.}

\begin{tcolorbox}[width=\columnwidth,colframe=blue!30,colback={blue!15},title={Mathematical 
    Core
 of 
 Quantum Algorithms},outer arc=0mm,colupper=black,colbacktitle=white,coltitle=blue!100] 
    In the \emph{quantum-circuit model}, quantum algorithms can be concisely formulated as applying a (possibly large) unitary matrix $U$ to an initial state $\ket{\psi_0}$ and performing a projective measurement w.r.t.\ an observable $M$ afterwards.
Often, the measurement has the purpose of estimating the quadratic form~\eqref{eq:meas_quad_form}.
In this case, the quantum algorithm outputs 
the expectation value 
\begin{align}\label{eq:quantum_algorithm_formula}
    \braket{\psi_0|U^\dagger MU|\psi_0}.
\end{align}\\[-11mm]
    \end{tcolorbox}

In the following, we summarize and discuss several fundamental 
quantum algorithms.

\textbf{Quantum Fourier Transform:}
The quantum Fourier transform (QFT) is one of the most important quantum algorithms, especially since it is frequently used as a subroutine (e.g., in Shor's algorithm).
To define it, we use the decimal representation of the computational basis states in~\eqref{eq:computational_basis_states} such that, e.g., $\ket6=\ket{110}$.
The QFT acts on a given initial state $\ket{\psi_0}$ by applying the discrete Fourier transform (DFT) on the probability amplitudes.
More precisely, for $k=0,\dots,2^n-1$, the computational basis state $\ket{k}$ is mapped to 
\begin{align}
    \frac{1}{2^{n/2}}\sum_{j=0}^{2^n-1}e^{2\pi i \frac{jk}{2^n}}\ket j.
\end{align}
The definition is extended to arbitrary states $\ket{\psi_0}$ via linearity.
As shown in more detail in~\cite[Section 5]{nielsen2011quantum}, the above-defined operation is, in fact, unitary and can be implemented based on applying suitable quantum gates.
At first glance, the QFT promises a remarkable speed-up over the classical DFT since it allows to Fourier transform $2^n$ values using only $\mathcal{O}(n^2)$ operations, whereas the DFT requires $\mathcal{O}(n2^n)$ operations.
The main bottleneck is that the probability amplitudes of the resulting output state cannot be accessed directly but only via the principles of measurement as outlined in Section~\ref{subsec:measurement}.

\textbf{Shor's Algorithm:}
Shor's algorithm allows to factorize integers into their prime factors in polynomial time.
Integer factorization forms the backbone of the RSA public-key cryptosystem, i.e., the security of this system relies on the fact that, currently, no algorithm is known to factorize integers in polynomial time on a classical computer.
Thus, a sufficiently large and reliable quantum computer could render RSA insecure.
Shor's algorithm consists of two main steps:
1) translating the factorization problem into a period finding problem and 2) solving the latter problem using the QFT, compare~\cite[Section 5]{nielsen2011quantum}, \cite{shor1997polynomial} for further details.

\textbf{Grover's Algorithm:}
On a classical computer, searching a single marked element in a list of overall $N$ elements takes $\mathcal{O}(N)$ operations.
Grover's algorithm~\cite[Section 5]{nielsen2011quantum}, \cite{grover1996fast} solves the same problem on a quantum computer in only $\mathcal{O}(\sqrt{N})$ operations, i.e., with a quadratic speedup.
While a quadratic speedup is less powerful than an exponential one as in Shor's algorithm, Grover's algorithm has still become one of the most popular quantum algorithms since it can be used to improve other quantum algorithms which are based on exhaustive search.

\textbf{Quantum Simulation:}
Simulating the time evolution of a quantum system is a fundamental problem in various domains such as drug design, quantum chemistry, or material science.
In fact, quantum simulation has inspired the original idea of a quantum computer by Richard Feynman~\cite{feynman1982simulating}.
Mathematically, quantum simulation boils down to solving the Schr\"odinger equation
\begin{align}\label{eq:schroedinger_equation}
\ket{\dot{\psi}}=-iH\ket{\psi}
\end{align}
for the quantum state $\ket{\psi}$ and the \emph{Hamiltonian} $H=H^\dagger$.
However, since the dimension of $\ket{\psi}$ scales exponentially with the number of qubits, quantum simulation is inherently challenging for classical computers.
In quantum simulation, the key idea is to decompose the Hamiltonian $H$ into a sum of Hamiltonians which only act non-trivially on a subset of the $n$ qubits, i.e., $H=\sum_k H_k$ for sufficiently sparse matrices $H_k=H_k^\dagger$.
The solution of~\eqref{eq:schroedinger_equation} can then be approximated via the Lie-Trotter formula as
\begin{align}
    \ket{\psi(t)}=e^{-iHt}\ket{\psi(0)}\approx\prod_k e^{-iH_kt}\ket{\psi(0)},
\end{align}
where each of the unitary matrices $e^{-iH_kt}$ can be implemented (approximately) via suitable quantum gates. 
While this is only an approximation since $e^{A+B}\neq e^Ae^B$ in general, the literature contains various results on bounding the resulting approximation error, see, e.g.,~\cite{lloyd1996universal,childs2021theory} and the references therein.

\subsection{Variational Quantum Algorithms}\label{subsec:vqas}

In the current NISQ era~\cite{preskill2018quantum,bharti2022noisy}, quantum devices are of limited size and significantly affected by noise.
VQAs~\cite{cerezo2021variational} have emerged as a promising class of algorithms to address these challenges.
The key idea is to design quantum algorithms that are smaller but adapt to potential inaccuracies in the hardware.
This is made possible by introducing parametrized quantum algorithms, where the individual unitary matrices depend on real-valued parameters.
To be precise, for a set of matrices $H_j=H_j^\dagger$ and parameters $\theta\in\mathbb{R}^N$, define 
\begin{align}
    U(\theta)=U_1(\theta_1)\cdots U_N(\theta_N)
\end{align} 
with each $U_j(\theta_j)=e^{-i\theta_jH_j}$ for $j=1,\dots,N$.
Inserting $U(\theta)$ into the equation~\eqref{eq:quantum_algorithm_formula} characterizing a generic quantum algorithm, we obtain a map $f:\mathbb{R}^N\to\mathbb{R}$ with
\begin{align}\label{eq:vqa_f_theta}
    f(\theta)=\braket{\psi_0|U(\theta)^\dagger M U(\theta)|\psi_0}
\end{align}
for some initial state $\ket{\psi_0}$ and observable $M=M^\dagger$.
In a VQA, the main idea is to iteratively adapt the parameters $\theta$ via a classical parameter update rule
\begin{align}\label{eq:vqa}
    \theta^+=g(\theta,f(\theta)),
\end{align}
in order to find a minimizer $\theta^*$ of the function $f(\theta)$.
Here, the map $g:\mathbb{R}^{N+1}\to\mathbb{R}^N$ is typically defined via some classical optimization routine, e.g., based on gradient descent~\cite{schuld2019gradient}. 
Note that executing~\eqref{eq:vqa} requires the execution of the quantum algorithm $f(\theta)$ as well as the classical computation $g(\theta,f(\theta))$, which is why VQAs are frequently referred to as hybrid quantum-classical algorithms.
From a control viewpoint, VQAs are especially interesting.

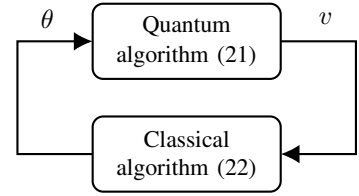
\begin{figure}
    \centerline{
        \begin{tikzpicture}[scale=1]
        \draw[-{Latex[length=2.5mm,width=2.5mm]},thick] (2,-1.5)--(1,-1.5)--(1,0)--(2,0);
        \draw[thick,rounded corners] (2,0.5) rectangle (4.5,-0.5);
        \node at (3.25,0.2) {\small Quantum};
        \node at (3.25,-0.2) {\small algorithm~\eqref{eq:vqa_f_theta}};
        \draw[-{Latex[length=2.5mm,width=2.5mm]},thick] (4.5,0)--(5.5,0)--(5.5,-1.5)--(4.5,-1.5);
        \draw[thick,rounded corners] (2,-1) rectangle (4.5,-2);
        \node at (3.25,-1.3) {\small Classical};
        \node at (3.25,-1.7) {\small algorithm~\eqref{eq:vqa}};
        \node at (1.4,0.3) {$\theta$};
        \node at (5.1,0.3) {$v$};
        \end{tikzpicture}%
    }
    \caption{Variational quantum algorithm (VQA) as feedback interconnection of the quantum algorithm~\eqref{eq:vqa_f_theta} (a static nonlinearity) and the classical algorithm~\eqref{eq:vqa} (a dynamical system).
}
    \label{fig:vqa}
\end{figure}

\begin{tcolorbox}[width=\columnwidth,colframe=blue!30,colback={blue!15},title={VQAs As Feedback Interconnections},outer arc=0mm,colupper=black,colbacktitle=white,coltitle=blue!100] 
    VQAs are feedback interconnections of a dynamical system (the classical algorithm) and a static nonlinear function (the quantum algorithm), compare Figure~\ref{fig:vqa}.
    Mathematically, this means 
\begin{equation}
    \theta^+=g(\theta,v)\quad\text{with}\quad
    v=f(\theta).
\end{equation}\\[-11mm]
    \end{tcolorbox}

While VQAs have received significant attention in recent years, they face a number of open challenges, in particular the design of the classical optimization scheme $g$ as well as the corresponding convergence analysis~\cite{cerezo2021variational}.
Given that they are, in fact, Lur'e systems, they lend themselves naturally to a theoretical analysis via systems-theoretic tools such as dissipativity~\cite{lessard2022analysis,scherer2022dissipativity}.

Let us conclude by mentioning several examples of VQAs.
The variational quantum eigensolver (VQE)~\cite{peruzzo2014variational} aims to determine the minimum eigenvalue of a matrix $H$, which is a challenging problem of central importance in various domains, in particular in quantum chemistry.
The quantum approximate optimization algorithm (QAOA)~\cite{farhi2014quantum}, on the other hand, tackles combinatorial optimization problems.
Given that many relevant problems arising in control involve integer optimization, the QAOA is an interesting candidate for using quantum computers in control, see~\cite{inoue2020model,deshpande2022quantum,schneider2024using}.
Finally, in variational quantum machine learning, the function $f$ in~\eqref{eq:vqa_f_theta} is employed as a function approximator for supervised learning~\cite{benedetti2019parameterized,schuld2020circuit,abbas2021power,jerbi2023quantum,berberich2023training}.

\subsection{Density Matrices}\label{subsec:density_matrices}

Density matrices are an alternative representation of quantum states that emerge naturally from the following thought experiment:
Suppose we are given an ensemble of quantum state vectors $\ket{\psi_i}$, $i=1,\dots,q$.
It is unknown in which of these states a given quantum system is, but we have a set of probabilities $p_i$, $i=1,\dots,q$, such that the system is in state $\ket{\psi_i}$ with probability $p_i$.
This information can be conveniently summarized via the \emph{density matrix}
\begin{align}\label{eq:density_matrix}
    \rho=\sum_{i=1}^qp_i\ket{\psi_i}\bra{\psi_i}.
\end{align}
The density matrix description of quantum states is equivalent (up to phases, cf.\ below)
to the state vector picture employed in the previous sections.
Density matrices can be mathematically more convenient to use, in particular in the context of quantum error correction~\cite{nielsen2011quantum} as well as in quantum systems and control theory, compare Section~\ref{sec:tosh}.
To provide a specific example, recall that multiplication of a quantum state by a global phase $e^{-i\varphi}$ does not affect the measurement outcome and therefore, strictly speaking, any quantum state $\ket{\psi}$ gives rise to an equivalence class $\{e^{-i\varphi}\ket{\psi}\mid\varphi\in\mathbb{R}\}$.
In the density matrix picture, on the other hand, global phase factors cancel out in forming the products
\begin{align}
    \ket{\psi_i}e^{i\varphi}e^{-i\varphi}\bra{\psi_i}=\ket{\psi_i}\bra{\psi_i}.
\end{align}
All operations defined in the previous sections can be equivalently formulated for density matrices.
The action of a quantum gate $U$ on $\rho$ can be easily computed by acting on each of the states $\ket{\psi_i}$ individually, i.e., 
\begin{align}
    \rho\mapsto \sum_{i=1}^qp_iU\ket{\psi_i}\bra{\psi_i}U^\dagger=U\rho U^\dagger.
\end{align}
Further, a projective measurement w.r.t.\ an observable $M=\sum_i\lambda_iP_i$ yields the outcome $\lambda_i$ with probability 
\begin{align}
    \mathrm{tr}(P_i\rho).
\end{align}
If the measurement returns $\lambda_i$, then the state collapses to 
\begin{align}
    \rho'=\frac{P_i\rho P_i}{\mathrm{tr}(P_i\rho)}.
\end{align}
It is simple to verify that, in the special case where $\rho$ represents one quantum state vector $\ket{\psi}$ with certainty, i.e., $\rho=\ket{\psi}\bra{\psi}$, the above formulas reduce to those presented in Section~\ref{subsec:measurement}.
States $\rho$ for which $\rho=\ket{\psi}\bra{\psi}$ holds with some $\ket{\psi}$ are called \emph{pure states}, whereas states as in~\eqref{eq:density_matrix} with $q>1$ are called \emph{mixed states}.
The trace of $\rho^2$ provides a simple criterion to distinguish between pure states ($\mathrm{tr}(\rho^2)=1$) and mixed states ($\mathrm{tr}(\rho^2)<1$). 
For single-qubit states $\rho\in\mathbb{C}^{2\times 2}$, this difference can be nicely illustrated via the Bloch sphere.
Here, pure states are all the points on the sphere whereas mixed states lie within the unit ball.
Intuitively, states that are closer to the center of the ball possess a higher level of uncertainty.
The state at the center has the maximum possible uncertainty and is, therefore, referred to as the \emph{maximally mixed} state.
Mathematically, it lies in equal superposition of the computational basis states $\ket0$ and $\ket1$ such that
\begin{align}
    \rho=\tfrac{1}{2}\ket0\bra0+\tfrac{1}{2}\ket1\bra1=\tfrac{1}{2}I_2.
\end{align}
%

%% file: part_TSH.tex


\newcommand{\hhH}{\widehat H}
\newcommand{\hhGA}{\widehat\Gamma}
\newcommand{\bG}{\mathbf{G}}
\newcommand{\fg}{\mathfrak{g}}
\newcommand{\fgl}{\mathfrak{gl}}
\newcommand{\su}{\mathfrak{su}}
\newcommand{\so}{\mathfrak{so}}
\newcommand{\bK}{\mathbf{K}}
\newcommand{\fk}{\mathfrak{k}}
\newcommand{\fp}{\mathfrak{p}}
\newcommand{\fm}{\mathfrak{m}}
\newcommand{\fn}{\mathfrak{n}}
\newcommand{\ketbra}[2]{\ensuremath{| #1 \rangle \langle #2 |}{}}
\newcommand{\adr}{\operatorname{ad}}
\newcommand{\Adr}{\operatorname{Ad}}
\newcommand{\reach}{\operatorname{Reach}}
\newcommand{\Mat}{\operatorname{Mat}}
\newcommand{\sspan}{\operatorname{span}}
\newcommand{\unity}{\ensuremath{{\rm 1 \negthickspace l}{}}}

\section{Exploiting a Unified {\sc Lie} Frame for Quantum Systems Theory and Control Engineering}\label{sec:tosh}
\noindent
The amount of oversimplification when reading this section under the 
overarching motto \/`{\em Break the symmetries of your (bilinear) quantum control system 
by choosing the right controls (or observables) in order to bring it under control
(or observation)\/!}\/' 
is---of course not quite negligible but---in no way criminal.
The reader is invited to bear with us when making the motto more
precise step by step in the sequel. In so doing, we 
wrap up Lie-theoretical intricacies in adapted matrix representations,
so the reader may keep the familiar \/`matrix times vector\/' picture.


\medskip
{\em Finite-dimensional} dynamical sytems pertinent to quantum-control engineering
of, e.g., $N$-level quantum gates
often boil down to the 
standard
form of {\em bilinear control systems} 
\cite{ControlHB,Sontag,Elliott09,DiHeGAMM08}
\begin{equation}\label{eq:bilinear}
\dot X(t) = -(A + \text{\large{$\Sigma$}}_j u_j(t) B_j) X(t)\quad\text{with}\quad X(0)=X_0\tag{$\Sigma$}
\end{equation}
as exemplified in the summarizing Tab.~\ref{tab:glossary}.
E.g., in a controlled Schr{\"o}dinger equation 
$ \ket{\dot{\psi}(t)} = -i (H_0 + \sum_j u_j(t)H_j) \ket{\psi(t)}$
$A,B$ are 
linear operators on a 
(finite-dimensional) Hilbert space $\mathcal H$ 
and $X(t)=\ket{\psi(t)}$ is a pure quantum state. 
In these closed systems,  
the non-switchable {\em drift term} $A$ is given by the system 
Hamiltonian $iH_0$, while the
$B_j$ are the {\em control Hamiltonians $iH_j$} governed by typically 
piece-wise constant control amplitudes
$u_j(\tau)\in\mathbb R$. In such a time interval, the time evolution
is $\ket{\psi(\tau)}=e^{-i\tau H}\ket{\psi(0)}=:U(\tau)\ket{\psi(0)}$.
Compiling quantum gates as in Section~\ref{subsec:gates} can be 
thought of (and done!) via solving an optimal-control problem for the system~\eqref{eq:bilinear} with variables as in Tab.~\ref{tab:glossary} (1').

In general and for open dissipative systems, states are described by 
{\em density operators} $\rho(t)=\sum_k p_k\ketbra{\psi_k}{\psi_k}$,
i.e.\ convex combinations of pure-state projectors 
as in 
Section~\ref{subsec:density_matrices}.
%
%
Then taking $X(t)=\ket{\rho(t)}$ as vectorised density operator, 
and conveniently moving to (super)operators in Liouville space
now with drift term $ i \hhH_0 + \hhGA$ 
(including possible relaxation $\hhGA $) {\em versus} controls 
as $u_j(t) i\hhH_j$ recovers the simple bilinear pattern of Eqn.~\eqref{eq:bilinear}. 
For the Hamiltonian part, $\widehat H \ket\rho$ corresponds to the commutator 
$[H,\rho]:= H\rho - \rho H$ also written as $\adr_H(\rho)$ in the sequel.

%
In the respective operator lift of a closed system, 
$X(t)$  takes the form
of a unitary conjugation map $\widehat U(t)$ 
acting on density operators $\ket\rho$ corresponding to $U(t)\rho U^\dagger(t)$
denoted $\Adr_{U(t)}(\rho)$---see the remark on tools and notations below. 

For deciding the feasibility of
quantum engineering tasks, in the sequel we  introduce
core definitions and systems-theoretical concepts for~\eqref{eq:bilinear} 
like controllability, reachabilty, observability etc. The concrete implementation can then be relegated, e.g.,
to numerical optimal-control algorithms as sketched in Sec.~\ref{sec:num-algs}.

{\footnotesize
\begin{table}[t]
\begin{center}
\caption{\mbox{Scenarios of Bilinear Quantum Control Systems \cite{OSID17}
}}
\label{tab:glossary}
\begin{tabular}{@{\hspace{1mm}}ll@{\hspace{4mm}}l@{\hspace{-1mm}}
r@{\hspace{1mm}}}
\\[-3mm]
\hline\hline\\[-2mm]
{Setting and Task }&\/`State\/'& Drift & Controls \\[0mm]
\multicolumn{1}{l}{$\dot X(t) = -\big(A + \sum_j u_j(t)  B_j\big) X(t)$} 
& \;$X(t)$ & $A$ &  $B_j$ \\[.5mm]
{} &          &      &  \\[-2mm]
\emph{closed systems:} &          &      &  \\[-.5mm]
\, (1\phantom{`}) pure-state transfer (fixed global phase) & \;$\ket{\psi(t)}$  & $i H_0$  & $i H_j$  \\[0mm]
\, (1') gate synthesis (fixed global phase) & \;$U(t)$ 
& $i H_0$  & $i H_j$  \\[0mm]
\, (2\phantom{`}) state transfer (modulo global phase) & \;$\ket{\rho(t)}$  & $i \hhH_0$  
& $i \widehat{H}_j$  \\[0mm]
\, (2') gate synthesis (modulo global phase) & \;$\widehat{U}(t)$
& $i \hhH_0$  & $i \hhH_j$  \\[2mm]
{} &          &      &  \\[-3.5mm]
\emph{open systems:}   &         &              &          \\[0mm]
\, (3\phantom{`}) state transfer I& \;$\ket{\rho(t)}$
& $i \hhH_0 {+} \hhGA$ &  $i \hhH_j$  \\[0mm]
\, (3') quantum-map synthesis I& \;$F(t)$
& $i \hhH_0 {+}  \hhGA$ &  $i \hhH_j$  \\[0mm]
\, (4\phantom{`}) state transfer II& \;$\ket{\rho(t)}$
& $i \hhH_0$ &  $i \hhH_j,\hhGA$  \\[0mm]
\, (4') quantum-map synthesis II& \;$F(t)$
& $i \hhH_0$ &  $i \hhH_j,\hhGA$  \\[1mm]
\hline\hline\\[-2mm]
\multicolumn{4}{l}{\scriptsize $\ket{\psi}$ is a pure-state vector, $\ket{\rho}$ a vectorised density operator, $\hhH$ is the Hamiltonian}\\[0mm]
\multicolumn{4}{l}{\scriptsize adjoint action 
generating unitary conjugation $\widehat U$ see notational remark below}\\[-6mm]
%
%
\end{tabular}
\end{center}
\end{table}
}
\subsection{Controllability and Reachability in Closed Systems}
\noindent
While a {\em linear control system}
$
\dot x(t) = Ax(t) +  B u\;\text{with}\; x_0=x(0)
$
is fully controllable 
\cite{Brockett_LinSys,Sontag}
provided its 
reachability matrix $[B, AB, A^2 B, \dots, A^{N-1} B]$ 
has full rank, controllability of a
{\em bilinear system} of the form Eqn.~\eqref{eq:bilinear} 
(henceforth on a compact connected Lie group $\bK$ with Lie algebra $\fk$, so $\bK={\langle\exp \fk\rangle}$) 
depends on its {\em system algebra}
\begin{equation}\label{eq:Lie-closure}
\fk_\Sigma :=\langle A, B_j \,|\,j=1,2,\dots,m\rangle_{\rm Lie}\;,
\end{equation}
where $\braket{\cdot}_{\sf Lie}$ denotes the {\em Lie closure} 
of taking the real span over nested commutators
until no new linearly indepedent elements are generated.
So the system algebra $\fk_\Sigma$ takes the form of a 
Lie algebra.\footnote{A {\em Lie algebra} $\fg$ (here over finite-dim.\ square matrices) is closed
under the commutator $[A,B]:=AB-BA$, and it is (i) bilinear in its arguments, (ii)
antisymmetric ($[A,B]=-[B,A]$) and (iii) satisfying  {\sc Jacobi}'s identity
$[A,[B,C]]+[B,[C,A]]+[C,[A,B]]=0$ $\forall A,B,C\in\fg$. NB: all square matrices
$\mathbb C^{N\times N}$ form the Lie algebra $\fgl(N,\mathbb C)$, while $\su(N)$
is a subalgebra
comprising its traceless skew-hermitian ($A^\dagger$=$-A$) part.}
It is the fingerprint of the dynamic system and it decides its controllability
as follows.

\medskip
\begin{tcolorbox}[width=\columnwidth,colframe=blue!30,colback={blue!15},title={Controllability of Bilinear Systems},outer arc=0mm,colupper=black,colbacktitle=white,coltitle=blue]    
   The bilinear control system~\eqref{eq:bilinear} is {\em controllable} on $\bK$, 
if its system Lie algebra $\fk_\Sigma$ of Eqn.~\eqref{eq:Lie-closure}
satisfies the  celebrated {\em Lie-algebra rank condition}
$\fk_\Sigma=\fk$ 
\cite{SJ72,JS72,Bro72,Bro73,Jurdjevic97}
and thus generates the entire Lie group $\bK$ by way of exponentiation 
$\bK={\langle\exp \fk_\Sigma\rangle}$.\\[-5mm]
\end{tcolorbox}   

\medskip
In closed quantum
systems (no dissipation) with $N$ levels (dimensions), first take $\fk_\Sigma$ as a subalgebra of $\su(N)$,
the latter comprising the $N\times N$ skew-hermitian (traceless) matrices. Each element can be
thought of as ($i$ times) a Hamiltonian direction `$i  H$' where $H$=$H^\dagger$ is
some Hamiltonian (i.e.\ `energy operator').
So---at first sight in the scenario (1 or 1') of Tab.~\ref{tab:glossary}---for a (closed) 
$N$\/-\/level quantum bilinear control system 
to be fully controllable on the pure-state vectors $\ket\psi$, it suffices that
$\bK={SU(N)}$, i.e.\ the system algebra comprises all the unitaries $\fk_\Sigma={\su(N)}$. 
For quantum information, this means that every unitary
quantum gate can be generated as $SU(N)={\exp \fk_\Sigma}$. Such a system
is 
{\em universal} 
as it allows to implement a universal quantum computer.

Recall that, by Eqn.~\eqref{eq:meas_quad_form}, a quantum mechanical expectation value of an observable $A$=$A^\dagger$
w.r.t. to a state $\ket\psi$ is of scalar-product form $\braket{A}_\psi = \bra\psi A\,\psi\rangle$, 
hence all states $\{ e^{i\phi}\ket\psi\,|\, \phi\in\mathbb R\}$ share the same
expectation value and are thus operationally indistinguishable.
Therefore, fixed {\em global} phases are artificial and do not correspond to any physically relevant
degree of freedom\,\footnote{in contrast to {\em relative} phases of one {\em{sub}}system w.r.t.\ other
subsystems as, e.g., in the Aharanov-Bohm experiment}. The same holds for global phases
of unitary time-evolutions $\{e^{i\alpha} U(t)\,|\, \alpha\in\mathbb R \}$.
Most conveniently, 
by projective construction they neither 
appear in the density operator $\rho(t)=\sum_k p_k\ketbra{\psi_k}{\psi_k}$
nor 
in its time evolution under unitary conjugation $\rho(t)=\Adr_{U(t)}(\rho_0)=U(t)\rho_0 U^\dagger(t)$.

That's why---see (2 or 2') of Tab.~\ref{tab:glossary}---it is 
natural to work with {\em density operators} $\rho$ and their time evolutions 
$\rho(t)$ henceforth as control problems on the {\em unitary orbit} $\mathcal O_U(\rho):=\{ U \rho_0 U^\dagger\,|\, U\in SU(N)\}$. 

\medskip
\noindent
{\em Tutorial Remark on Tailored Toolbox and Notations:}\\%
To solve matrix equations as $[H,X]=HX-XH=0$ for $X$, it
is convenient to vectorise the matrix $X$ to the vector $\ket X$ 
defined as the stacked column
of all columns of $X$. In this convention, one gets the useful formula 
$\ket{AXB}=(B^\top\otimes A)\ket X $, see Chp.~4. of \cite{HJ2}.
Then the solution sets to $[H,X]=0$ and
$(\unity\otimes H - H^\top\otimes\unity)\ket X = \ket 0$ interrelate, as
the kernel of $\widehat H:=(\unity\otimes H - H^\top\otimes\unity)$
represents the {\em commutant} of $H$ in vectorised form. Likewise,
the {\em joint commutant} of a set $\{H_j\}$ with, e.g., 
$j=1,2$ can conveniently be read from
the joint kernel of the stacked matrix $\left[\begin{smallmatrix} \widehat H_1\\\widehat H_2\end{smallmatrix}\right]$.
 
 With $H$ hermitian, 
$i\widehat H$ may be thought of as representing $i\adr_H$.
Therefore, $\exp( -it\adr_H) = \Adr_{U(t)}= U(t)(.)U^\dagger(t)$ (where $U(t):=e^{-itH}$)
 can be represented in the vectorised picture by $\widehat U(t)=\exp(-it\widehat H)=\bar U\otimes U$.  
 For staying in the familiar \/`matrix times vector\/' representation, think of
 $\Adr_U(\rho)$ as $\widehat U\ket\rho$ and of $\adr_H(\rho)$ as $\widehat H\ket\rho$.

 A set of square matrices like $\{iH_j\}_{j=0}^m$ is called {\em irreducible} if and only if,
 as linear operators, they have no common non-trivial invariant linear subspaces
 (other than the full space or the zero-space). This is guaranteed if  
 their {\em joint commutant is trivial} (i.e. just consists of multiples of the identity), 
 since then there is no common reducing projector.
 
 A key to the following symmetry considerations is that irreducibility of $\{iH_j\}_{j=0}^m$  does
{\em not} necessarily imply irreducibility\footnote{In the adjoint representation the 2 trivial symmetries are $\unity^{\otimes 2}$
and $\ketbra\unity\unity$.} 
of $\{i\widehat H_j\}_{j=0}^m$, 
unless
the $\{i H_j\}_{j=0}^m$ {\em generate the entire Lie algebra} $\su(N)$
by $\langle i H_j \,|\,j=0,1,2,\dots,m\rangle_{\rm Lie}$, which means
{\em the corresponding bilinear control system is fully controllable} \cite{ZS11}.

\medskip
Now we have the proper operational setting for the following controllability condition.
\medskip
\begin{tcolorbox}[width=\columnwidth,colframe=blue!30,colback={blue!15},title={Full Controllability (in Closed Quantum Systems)},outer arc=0mm,colupper=black,colbacktitle=white,coltitle=blue]    
A closed $N$\/-\/level {\em quantum} bilinear control system 
(also comprising mixed states)
is fully controllable on density operators $\rho$ 
if and only if
its system algebra is $\fk_\Sigma=\adr_{\su(N)}$. It also generates the entire
group of unitary conjugation $\bK_\Sigma=\Adr_{SU(N)}$ by $\exp\fk_\Sigma=\Adr_{SU(N)}=\{U(.)U^\dagger\,|\, U\in SU(N)\}$
and thus all unitary gates.\\[-5mm]
\end{tcolorbox} 



As an immediate consequence, in fully controllable systems one has the following reachable set of states $\rho$.

\begin{tcolorbox}[width=\columnwidth,colframe=blue!30,colback={blue!15},title={Reachable Sets in Fully Controllable Systems},outer arc=0mm,colupper=black,colbacktitle=white,coltitle=blue]    
The reachable set to an initial state 
$\rho_0$ of a {\em fully controllable} closed quantum bilinear control system ($\Sigma$)
with system algebra $\fk_\Sigma=\adr_{\su(N)}$
 corresponds to the entire unitary orbit 
$\reach_\Sigma(\rho_0)=\mathcal O_U(\rho_0)=\{ U \rho_0 U^\dagger\,|\, U\in SU(N)\}$.\\
It thus equals the {\em isospectral set of $\rho_0$}
defined as the set of 
all density operators $\rho$ sharing the same eigenvalues with the
initial state $\rho_0$.\\[-5mm]
\end{tcolorbox}

Shifting back to quantum gates
one arrives at the following result in the general (not necessarily fully controllable) case.

\begin{tcolorbox}[width=\columnwidth,colframe=blue!30,colback={blue!15},title={Reachable Gates},outer arc=0mm,colupper=black,colbacktitle=white,coltitle=blue]    
The reachable set of gates $\{K(.)K^\dagger\}$
of a closed quantum bilinear control system ($\Sigma$)
with system algebra $\fk_{\Sigma}$
is given by $\bK_{\Sigma}:=\exp\fk_{\Sigma}$.\\[-5mm]
\end{tcolorbox}

Accordingly one gets
the following general reachability result for quantum states $\rho$.

\begin{tcolorbox}[width=\columnwidth,colframe=blue!30,colback={blue!15},title={Reachable Sets},outer arc=0mm,colupper=black,colbacktitle=white,coltitle=blue]    
The reachable set to an initial state $\rho_0$
of a closed quantum bilinear control system ($\Sigma$)
with system algebra $\fk_{\Sigma}$
 is given by the orbit
$\reach_{\Sigma}(\rho_0)=
\{K \rho K^\dagger\,|\, K(.)K^\dagger \in \bK_{\Sigma}\}$
determined by the system algebra via $\bK_{\Sigma}=\exp\fk_{\Sigma}$.\\[-5mm]
\end{tcolorbox}
\noindent
Thus in closed systems a reachable set of states $\reach_{\Sigma}(\rho_0)$ is always
a subset of the full unitary orbit $\mathcal O_U(\rho_0)$. 

\bigskip

\medskip

\noindent
{\em Application: Can Specific Gates be Engineered?}\\
For quantum engineering of unitary gates, the following result is of practical relevance.
Even if a quantum system ($\Sigma$) is {\em not fully controllable} so that it does not allow for implementing
all unitary quantum gates, it may still serve to implement specific gates. And,
e.g., on a dedicated quantum chip it may even do so in
a faster or more robust fashion than a fully controllable universal setup. 
By way of the system algebra $\fk_\Sigma$, the system theoretic setting here 
allows for the following straightforward feasibility check:

Let $\hat L_t$ be a (negative) matrix logarithm of a desired target unitary gate $\hat U_t$
(in adjoint representation), 
i.e. $\hat U_t=e^{-\hat L_t}$.
Then it can be realised
by the control system ($\Sigma$) if $\hat L_t$ can be spanned by the matrices in its system algebra $\fk_\Sigma$,
which in turn can be checked by vectorising the matrices to test whether 
$\rank(\fk_\Sigma)=\rank(\fk_\Sigma, \hat L_t)$. 
---
This is an example of how in our Lie frame
of quantum systems theory, engineering questions can often be broken down
to simple linear algebra.

\bigskip
\subsection{Symmetry Assessment of Controllability and Simulability}
\noindent
Although the system algebra $\fk_\Sigma$ is a highly useful
fingerprint encapsulating the capabilities of the parent bilinear
control system ($\Sigma$),  it may be tedious to come by via calculating
the entire Lie closure of Eqn.~\eqref{eq:Lie-closure}.

So, how can we simplify the assessment?
Often the properties of a control system may already be read from its {\em symmetries},
which mirror invariances under controls. It is intuitively obvious that a
control system with non-trivial symmetries comes with invariant \/`constants of the motion'\/
that preclude full controllability. It just takes a few steps to make these symmetry
arguments rigorous:

\medskip
\noindent
{\em Definition: Dynamic Symmetries of Bilinear Systems}\\
Let ($\Sigma$) be an $N$-level bilinear control system with system algebra 
$\fk_\Sigma=\braket{i\adr_{H_j}\,|\,j\,=\,0,\dots, m}_{\sf Lie}$, where the adjoint representation is crucial.
Then its {\em symmetries} are defined by the commutant of $\fk_\Sigma$, i.e.\
the set $S_\Sigma$ of all matrices simultaneously commuting with the 
entire system algebra.

Now the beauty of Lie theory comes in handy:
Since  
$\fk_\Sigma$
carries the structure of a Lie algebra, 
one simply\footnote{i.e.\ by {\sc Jacobi}'s identity}  gets its
entire symmetries just by the symmetries of the few(!) drift and control terms 
(as generators of the system Lie algebra) by
$S_\Sigma:=\{S\in \Mat(N^2)\,|\, [S,i\adr_{H_j}]=0 \;\forall\; j=0,1,\dots,m\}$\footnote{ {\sc Jacobi}'s identity likewise endows the symmetries $S_\Sigma$ themselves 
with the structure of a Lie algebra.}.

\begin{tcolorbox}[width=\columnwidth,colframe=blue!30,colback={blue!15},title={Trivial Symmetry Guarantees Controllability},outer arc=0mm,colupper=black,colbacktitle=white,coltitle=blue]    
A closed 
{\em quantum} bilinear control system 
is fully controllable on density operators $\rho$ 
(or on gates of unitary conjugation) if and only if
its system algebra $\fk_\Sigma$ just shows trivial symmetries,
i.e.\ $\dim(S_\Sigma)=2$.\\[-5mm]
\end{tcolorbox} 

The core insight can be made lucid by a standard group-theory 
argument\footnote{although the original rigorous proof of this fact was rather involved
\cite{ZS11} in relating back to {\sc Dynkin}'s work~\cite{Dynkin57} reproduced in~\cite{Dynkin2000} 
}: 
if the system algebra $\fk_\Sigma$ (adjoint representation) is irreducible in the sense of trivial symmetries, 
then there is no non-trivial
invariant subspace supporting control-invariant states $\rho$
other than multiples of the identity. 

The symmetry assessment is very powerful and readily extends to quantum simulation.
In quantum simulation one picks a well controllable system ($\Sigma_a$)
realized in an experimentally  handable setup in order to simulate the dynamics of
another system ($\Sigma_b$) of interest similar in dynamics, which, however, is 
difficult to control experimentally, cp. Sec.~\ref{subsec:algorithms}.

\medskip
\noindent
{\em Definition: Simulability in Bilinear Control Systems}\\
A bilinear quantum control system ($\Sigma_a$) with system algebra $\fk_{\Sigma_a}$
can {\em simulate} another one ($\Sigma_b$) with system algebra $\fk_{\Sigma_b}$ if
the simulating dynamics encapsulate the simulated dynamics, i.e.\
if and only if (up to isomorphism) $\fk_{\Sigma_b}\subseteq\fk_{\Sigma_a}$~\cite{ZS11}.
\medskip

Again, the explicit system algebras $\fk_{\Sigma_{a,b}}$ 
may be tedious to come by via the corresponding Lie closure of Eqn.~\eqref{eq:Lie-closure}.
Now let $S_{\Sigma_a}$ and $S_{\Sigma_b}$ denote the respective symmetries of the
simulating and the simulated quantum bilinear control system.
The powerful intuitive symmetry argument that the {\em simulated} system may not break the
symmetries of the {\em simulating} system can be 
expressed as a necessary condition $S_{\Sigma_b}\supseteq S_{\Sigma_a}$, which was
generalised to arbitrary compact cases and made precise in 
\cite{ZZS+15} 
by also taking into account the dimensions of central projections (adding up to the necessary
condition above the sufficient condition)
not discussed here.

\bigskip
\subsection{Symmetry Assessment of Observability, Accessibility,
 and Tomografiability}
\noindent
Recall that 
a {\em linear control system}
$
\dot x(t) = Ax(t) +  B u\;\text{with}\; x_0=x(0)\;\text{and observed by\;}
y(t)=C x(t)
$
is {\em observable} 
if its 
observability matrix $[C, AC, A^2 C, \dots, A^{N-1} C]^\top$ 
has full rank.
On the other hand, observability of 
the {\em bilinear control system}~\eqref{eq:bilinear}\footnote{again on a compact connected Lie group $\bK$ with Lie algebra $\fk$}
{\em in quantum dynamics} comes
with an observation term in form of a 
scalar product for the expectation value of $C$ w.r.t. the state $X(t)$
as $Y(t)=\tr\{C X(t)\}$ brought about
by an {\em observable} $C$=$C^\dagger$.
In accordance with scenario (2) of Tab.~\ref{tab:glossary}, it 
pays to represent the observation term in vectorised
form in line with the control system henceforth {\em on the orbit(!)} as
\begin{eqnarray*}\label{eq:bilinear-obs}
\ket{\dot X(t)} &=& -(\widehat A + \text{{\large{$\Sigma$}}}_j u_j \widehat B_j) \ket{X(t)}\;\text{with}\; \ket{X(0)}=\ket{X_0}\\[-0mm]
Y(t) &=&\braket{C \,|\,X(t)}\,. \hspace{45mm} \text{($\Sigma'$)}
\end{eqnarray*}

\noindent
In this setting one gets the following modifications of \cite{dAle03,dAless22}.

\medskip
\noindent
{\em Definition: Observability Space of Quantum Bilinear Systems}\\
Let ($\Sigma'$) be an $N$-level bilinear control system 
observed by $C$ and  with system algebra
$\fk_{\Sigma'}$ as in Eqn.~\eqref{eq:Lie-closure}. 
Denote the traceless part\footnote{$\tilde{C}:=C-\tfrac{\tr C }{N}\unity_N$} 
of $C$ as $\tilde C$ and
let $k^{(\nu)}(\tilde C):=k_1 k_2 \cdots k_\nu (\tilde C)$ be the set 
of all products of adjoint actions from the system algebra
$k_i\in\fk_{\Sigma'}$ ($i=1,\dots,\nu$) with $\tilde C$ of order $\nu$.
Then the {\em observability space} of 
($\Sigma'$)
w.r.t. the observable $C$
is defined 
\begin{equation}\label{eq:obs-space}
\mathcal O_{\Sigma'}(C):= \sspan_{\mathbb R} \big\{
{\tilde C},
k^{(1)} ({\tilde C})  ,
k^{(2)}  ({\tilde C}),
\dots ,
 k^{(N^2-1)} ({\tilde C})
\big\}\,.\vspace{2mm}
\end{equation}


\noindent
{\em Definition: Observability in Quantum Bilinear Systems}\\
Let ($\Sigma'$) be an observed $N$-level bilinear control system. 
Then ($\Sigma'$)  {\em is observable by $C$}
if for any pair of states $X_1, X_2$
the equality of their expectation values
$\tr\{C X_1(t)\} = \tr\{C X_2(t)\}$ for all $t\in\mathbb R$ and joint controls $u_j(t)$
implies the equality of the states $X_1 = X_2$.
---
This is the case if and only if 
$\mathcal O_{\Sigma'}(C)$
comprises {\em all hermitian matrices}  \cite{dAle03,dAless22}, 
as such an observability space 
is 
{\em informationally complete}. 

Note that a fully controllable bilinear control system
is always observable as well by any non-trivial $C$, which gives a sufficient condition that in turn 
is not necessary:
systems that fall short of full control may still be observable w.r.t. a specific observable $C_1$,
while for other observables $C_2$ they are not---see the worked example below.

Again, it may be tedious to calculate the entire observability space of Eqn.~\eqref{eq:obs-space}.
Hence yet again symmetry assessment helps: Intuition suggests that the dynamic symmetries $S_{\Sigma'}$
of an observed bilinear quantum control system ($\Sigma'$) with system algebra $\fk_{\Sigma'}$ (if any) have to
be broken by the observable $C$.
The following makes this argument rigorous (details in \cite{WSH25}).

\begin{tcolorbox}[width=\columnwidth,colframe=blue!30,colback={blue!15},title={Trivial Symmetry Guarantees Observability},outer arc=0mm,colupper=black,colbacktitle=white,coltitle=blue]    
A closed 
{\em quantum} bilinear control system ($\Sigma'$) with system algebra $\fk_{\Sigma'}$
is {\em observable} by $C$ if and only if for its projector $P_C:=\ketbra{\tilde C}{\tilde C}$
the joint commutant to $\fk_{\Sigma'}$ and $P_C$ is again two-dimensional.\\[-5mm]
\end{tcolorbox}

\medskip

\noindent
{\bf Worked Example:}\\[-6mm] 
\begin{figure}[h]
\begin{center}
        \includegraphics[width=0.75\columnwidth]{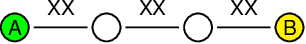}
\end{center}
\caption{$XX$-coupled 4-qubit chain with local controls on the ends.}
\label{fig:so10}
\end{figure}\\
\noindent
Take a linear chain consisting of four spin-1/2  qubits.
In the basis of Pauli operators written as strings\footnote{so $1xx1$ stands for 
$\unity\otimes\sigma_x\otimes\sigma_x\otimes\unity$}, 
let the drift term be what the
physicists call nearest-neighbour $XX$-coupling, i.e.\ 
$H_0:= (xx11+yy11+1xx1+1yy1+11xx+11yy)$. The
local control Hamiltonians  $H_1:=x111, H_2:=y111$ acting as
$x$- and $y$-controls on qubit (A) and $H_3:=111x, H_4:=111y$
as controls on qubit (B) make the coloured ends in Fig.~\ref{fig:so10}
controllable, while the white qubits are uncontrolled.


This system has non-trivial symmetries\footnote{The commutant to its system algebra $\fk_\Sigma$ in adjoint representation is 3-dimensional and the system algebra
is (the adjoint representation of) the orthogonal algebra $\so(10)$ embedded in the one
for the four-qubit unitary algebra $\su(16)$ as elaborated on in~\cite{ZS11}.}.
When choosing the observable to be $C_1:=xxx1$, the symmetries are preserved and thus
the system fails to be 
observable\footnote{The observability space of $C_1$ under $\fk_\Sigma$ has just 210 dimensions,
being the orthocomplement to $\so(10)$ in $\su(16)$, while for $C_2$ it
exhausts all the 255 dimensions of traceless (skew)hermitian $16\times 16$ matrices as in $\su(16)$.}
because it is taken from a $\fk_{\Sigma'}$-invariant subspace. The
proper choice in order to get full observability is, e.g., to make it $C_2:=xxx1 + 1z11$,
which breaks the symmetry of $\fk_{\Sigma'}$-invariant subspaces---as corroborated by trivial joint commutants.

[{\em For the expert:}
in $\fg=\su(N)$ we take the orthogonal decomposition $\fg=\fk\oplus\fm$
with $\adr_\fk(\fk)\subseteq \fk$ and 
$\adr_\fk(\fm)\subseteq \fm$, where (by mild abuse of language)
we identify the system algebra $\fk_{\Sigma'}$ in its adjoint representation with $\adr_\fk$
to see $\fk$ and $\fm$ as $\fk_{\Sigma'}$-invariant subspaces.

\begin{tcolorbox}[width=\columnwidth,colframe=blue!30,colback={blue!15},title={Observability Limited to Symmetry Sectors},outer arc=0mm,colupper=black,colbacktitle=white,coltitle=blue]    
A closed quantum bilinear control system ($\Sigma'$) with system algebra $\fk_{\Sigma'}$
is {\em observable on the smallest $\fk_{\Sigma'}$-invariant subspace $\fn\subset\fg$ supporting $\tilde C$}  
when restricting observability to those $A_\nu$ whose 
(non-trivial) $\fk_{\Sigma'}$-invariant support is included in $\fn$.
This requires the joint commutant of $\fk_{\Sigma'}\; \text{and}\;{P_{\tilde C}}$ 
to contain a (non-trivial)
projector  
$P_\fn$ so that
$$P_\fn\ket{\tilde C}=\ket{\tilde C}\; (\text{besides}\; P_\fn\ket{\tilde A_\nu}=\ket{\tilde A_\nu}).$$
 \vspace{-7mm}
\end{tcolorbox}
\noindent
NB: Full observability
is recovered in the \/`trivial'\/ case of $P_\fn=\unity^{\otimes 2}-\ketbra\unity\unity$ 
i.e.\ a two-dimensional commutant (s.a.).]

\medskip
Next, let us address accessibility,
which in open systems is strictly weaker than full controllability: 
A quantum bilinear control system
($\Sigma$) is {\em accessible} from a given initial state $X_0$, if the reachable set
$\reach_\Sigma(X_0)$ has non-empty interior. Roughly speaking \/`it has to touch all dimensions\/'.
In closed systems the reachable set of states takes the form of a $\bK$-orbit
as above. 
While in closed systems
controllability and accessibility coincide, 
in {\em open Markovian} systems (where controllability is lost),
symmetry arguments may be used to decide accessibility as sketched in the outlook.
 
Combining these notions suggests that given an initial state $X_0$, its dynamics under
a quantum bilinear control system ($\Sigma'$) is (fully) {\em tomografiable} 
by an observable $C$, if the system is both {\em accessible} from $X_0$ and (fully) {\em observable} by $C$. 
The respective weakening to symmetry sectors of non-trivial $\fk_{\Sigma'}$-invariant support 
as above is obvious.

Tomografiability would also lend itself as a feasibility criterion for Kalman filtering.

\medskip

\subsection{Outlook: Reachability in Open Quantum Systems}\label{sec:reach-open}
\noindent
Open Markovian~\cite{Wolf08a,DHKS08}
quantum bilinear control systems can be treated similarly by scenarios 
(3, 3') and (4, 4') in Tab.~\ref{tab:glossary}\footnote{if $\Gamma$ in Tab.~\ref{tab:glossary} is of Lindblad form~\cite{Koss72,GKS76,Lindblad76}}. 
In \mbox{$N$-level} systems, the system algebra $\fg_\Sigma$ 
then becomes a subalgebra of the $N^2\times N^2$ matrices $\fgl(N^2,\mathbb R)$. And the reachable sets change from
Lie-group orbits (as in the closed systems above) to Lie-{\em semigroup} orbits
as shown in~\cite{DHKS08} by 1:1 correspondence between Lie and Markov
properties in
the rich structure of Markovian quantum systems (sparked by the seminal
works around Kossakowski and Lindblad~\cite{Koss72,GKS76,Lindblad76}).

Depending on the sets of fixed points in such open systems, there are two families: 
in the {\em unital} case, the maximally mixed state $\rho=\tfrac{1}{N}\unity$ counts
among the fixed points, whereas in the {\em non-unital} case it does not. 

Motivated by earlier results~\cite{OSID17}, the {\em symmetry assessment} of reachability
elaborated on above
seems to extend to {\em Markovian} open bilinear quantum control systems of scenarios (3',4'): 
ongoing research reveals that \mbox{$n$-qubit}
unital bilinear control systems with a system algebra $\fg_\Sigma$ the symmetries to which are just
two-dimensional\footnote{and coincide with the trivial ones in the closed systems above} are (map)accessible, while in non-unital $n$-qubit bilinear control systems, (map)accessibility 
comes with one-dimensional trivial symmetry of the system algebra $\fg_\Sigma$.

\bigskip

Thus engineering questions of controllability, reachability, accessibility, observability, and tomografiability
can be answered in the {\em emerging unified Lie frame of quantum systems theory}. The doability
(existence) problems often
break down to decision or membership problems in linear algebra related to the respetive
dynamic system Lie algebras.   

Separating these existence questions from constructive proofs makes the methods more general and
more versatile. On the other hand, it invites {\em numerical methods} to address optimal-control
engineering in the specific scenario of an experiment taking into account its parameters
and constraints without need of (over)simplification or approximation making paper-and-pen
constructions viable.

\subsection{Implementations: Numerical Quantum Optimal Control}\label{sec:num-algs}
\noindent
For devising control pulses for experimental quantum setups,
numerical optimal-control methods play a key role see, e.g., 
the European Roadmap to Quantum Control Engineering~\cite{Koch22}.
Here variants of gradient-assisted methods like {\sc grape}~\cite{KG04b}
(incl.\ quasi-Newton
extensions \cite{PRA11}) are
established as widely applicable and popular.
They embrace the important {\sc Prontryagin} Maximum Principle~\cite{Pont64},
which comes into play whenever running costs are taken into account~\cite{KG04b}.
For robustness in numerical optimal quantum control and its roots in {\em ensemble controllability} also see \cite{Braun14}, Chp. 4.3 in~\cite{DongPetersen2023},
and \cite{likhaneja2009,beauchard2010controllability,dirr2012,Turinici15}, respectively
as well as Sec.~IV. 

\subsection{Exploiting Thermal Resources}
\noindent
With numerical methods at hand, one can easily go beyond the scenarios (3 and 3') in
Tab.~\ref{tab:glossary} and even use (Markovian) noise terms ($\Gamma$) as additional controls
(4 and 4').
Experimentally, this is made possible in the context of
superconducting qubits ({\sc Gmon}s) coupled---in a switchable way(!)---to an open
transmission line as, e.g.,  in \cite{Mart14,McDermott_TunDissip_2019}, 
among them a former group at Google.

Coupling a (coherently fully controllable) quantum bilinear system in a switchable way to 
the open transmission line acts 
when seen from within the quantum system\footnote{i.e.\ by projection onto the system} 
as standard amplitude-damping noise~\cite{nielsen2011quantum}, now
time-modulated as additional control. In the language of quantum thermodynamical
resources it serves as a switchable coupling to a temperature-zero bath.
This is an extremely powerful resource, because any $N$-level coherently fully
controllable system coupled via (at least) one qubit to such a bath acts transitively
on the set of now mixed quantum states (density operators $\rho$): it can
transform any given initial state $\rho_0$ in to {\em any} desired target state $\rho_T$
in the sense that the (closure\footnote{since one may have to wait for thermal exponential decay} of the) 
reachable set  to any initial state
exhausts the entire set of density operators on that Hilbert space.
So  $\overline{\reach_{\Sigma_T}(\rho_0)}=D(\mathcal H)$
as has been shown in \cite{BSH16} for the concrete parameter setting 
of \cite{Mart14}---in accordance with the symmetry considerations of Sec.~\ref{sec:reach-open}. 

These findings have also initiated to exploit the limits of Markovian thermal relaxation
seen as an additional resource in a more fundamental way~\cite{OSID23,SICON24}. 

{\em Conclusion:}
As to symmetries in systems theory, the
initial motto is a {\em natural guideline} to be taken over to numerical
approaches. To get a quantum system with drift term $A$ under control,
the more the control terms $B_j$ (Hamiltonians) 
can be chosen such as to {\em break the symmetries of the drift term}, the more control
one gets over the system. This has been elucidated, e.g., for controlling atoms in a
cavity in an ({\em infinite-dimensional}) Jaynes-Cummings model, where the control terms on the atom break
the symmetry of the oscillator~\cite{KZSH}. The same concept can be readily extended to couple
a system of an atom and cavity to a mechanical oscillator for obtaing non-classical
mechanical states with high accuracy~\cite{bergholmQST2019} 
using {\sc grape}-based numerical optimal control in a concrete experimental setup.

%% file: part_RK.tex

\section{Robust Synthesis of Quantum Gates}
\label{sec:robust_quantum_control}

\subsection{Robustness}

As explained in Section~\ref{sec:quantum_algorithms}, quantum
algorithms consist of a combination of qubits, quantum gates, and
measurements.  In experimental realizations of quantum computers,
perturbations due to noise pose a key challenge as they may affect the
(ideal) outcome of the computation to an extent that the overhead
required for error correction becomes out of reach in the near-term,
\eg, \cite{Preskill40:2023}.  To address this problem, robustness is
key, meaning that, despite imperfections or perturbations, each
\emph{actual} gate operation should be as close as possible to its
ideal.  In this section, we introduce a recently proposed robust
quantum control approach~\cite{KosutBR:2022} which borrows ideas from
the control community by modeling Hamiltonian uncertainty using a
set-membership approach, \eg,
\cite{Zames:1966,DesVid:1975,id4c:92,KosutLB:92,LMI94,ZhouDG:96}.  We
also establish a limit of robust performance if certain conditions are
met. This is not \emph{the} limit, that is not known.

There is much written on the subject of perturbed systems. Of special
note are the classic works of Bellman, Coppel, and Hale
\cite{Bellman:53,Coppel:1967,Hale:80}.  These works show how to
transform certain classes of systems to a standard perturbation form,
and secondly, show how system properties are maintained as long as the
perturbation has a small effect in a well-defined sense. Similar theory
follows for robustness of feedback systems via the \emph{Small-Gain
  Theorem}, \eg, \cite{Zames:1966,DesVid:1975}, and generalizations
for multiple and various uncertainty models, \eg,
\cite{LMI94,ZhouDG:96}.

Expanding upon the \emph{Method of Averaging} in \cite{Hale:80} we
modify the standard ``averaging transformation'' to be suitable for
quantum systems.  This resulting theory can be viewed through the lens
of the Small Gain Theorem as well as an optimization-based approach
whose underlying mechanism is supported by several known control
design methods similar to dynamical decoupling, \eg, see
\cite{ViolaKL:99,Uhrig:07,DDPRA:2011,GreenETAL:2013,HaasHamEngr:2019,QCTRL:2021}
and the references therein.  These approaches, and the one presented
here, differ from the typical quantum optimal control formulation
where the objective is to minimize fidelity error. In contrast, the
optimization problem here is posed with \emph{two} objectives:
fidelity error along with a term dependent on the size of an
interaction representation derived from the characteristics of the
sources of uncertainty. In the simplest case, this
\emph{regularization} part of the objective function is the first term
in the Magnus expansion of a transformed system unitary: \emph{the
  interaction unitary.}

  \subsection{Optimization} 
  \label{sec:optQC}
  
  The robust quantum control objective is typically formulated as an
  optimization problem to maximize a worst-case or average case fidelity (compare Section~\ref{sec:fid})
  with respect to a set of uncertainties represented by bounded
  sets, and subject to constraints on the optimization variables. For
  model-based control optimization, although the details vary, and are
  very important, most problems fit the following formulation:
  \beq[eq:Fgen]
  \bea{ll}
  \mbox{maximize}&
  \left\{
  \bea{ll}
  \mbox{worst-case}  & F=\min_\del\ F(\th,\del) 
  \\
  \mbox{or}
  \\
  \mbox{average-case} & F=\langle F(\th,\del)\rangle_\del 
  \eea
  \right.
  \\
  \mbox{subject to}  
  &
  \theta\in\Theta,\ \delta\in\Delta
  \eea
  \eeq
  Here, $F$ denotes the fidelity~\cite{nielsen2011quantum}, which quantifies the overlap between a given to-be-realized quantum gate and the actually realized, imperfect quantum gate, i.e., the deviation of $F$ from $1$ quantifies their difference, compare~\eqref{eq:UJfid}.
  The optimization variables $\theta$ typically include control
  amplitudes and phases produced by a field generator, or commands to
  the field generator if generator dynamics are of concern. Other design
  variables can include parameters corresponding to different
  circuit/gate layouts, material properties, etc. The design variables
  typically reside in a convex set $\Theta$. The uncertain variables
  $\delta$ are in a set $\Delta$ which may not be convex, though often
  well represented as such depending upon the specific physical
  implementation, i.e., the uncertainty set can be combinations of
  deterministic and random variables. Even if all the constraints are
  convex, the bilinear control nature of the quantum evolution means
  that the control problem in any form is not a convex optimization
  problem. As a result all solutions are iterative.
  
  Fortunately, however, the control landscape topology is favorable,
  \ie, almost always no traps, only saddles \cite{RabitzHR:04}. One
  route is via \emph{sequential convex programming}
  (SCP)~\cite{KosutGB:13} where robust performance can be attained in
  the presence of a list of inequalities associated with each
  uncertainty set. In addition to control parameters, the optimization
  can be used to select any physical adjustable parameters such as
  Hamiltonian couplings, material parameters, geometry of the qubit
  layout (an important consideration for QEC involving sparsely
  connected processor architectures), etc. For ``small'' deterministic
  and/or probabilistic perturbations, local approximations by convex
  functions will suffice rather than lists of inequalities.
  
  A related route to robust quantum control which can handle
  uncertainties more efficiently is to view the system through the lens
  of an \emph{interaction representation}, e.g., the basis for filter
  function formulations
  \cite{GreenETAL:2013,Kaby:2014,QCTRL:2021,HaasHamEngr:2019,Le:2022},
  our recent approach based on the method of averaging \cite{KosutBR:2022},
  and dynamical decoupling (see below)

\subsection{Nominal and perturbed system}

To illustrate the main ideas we assume a quantum system of dimension
$n$ whose {\em nominal} (uncertainty-free) dynamics produces the
unitary and state evolution for $t\in[0,T]$ given by
\beq[eq:Unom]
\beasep{1.25}{l}
i{\dot U}_{\mathrm{nom}}(t) = \Hb(t)\Unom(t), \quad \Unom(0)=I_n
\\
\ket{\psi_{\rm nom}(t)} = \Unom(t) \ket{\psi_{\rm nom}(0)}
\eea
\eeq
These dynamics correspond to the bilinear control system~\eqref{eq:bilinear} with $X=\Unom$ and $\Hb=A+\sum_ju_jB_j$ comprising both the drift as well as the control terms (compare scenario (1') in Tab.~\ref{tab:glossary}).
Similarly the {\em perturbed} (uncertain) system evolves according to
\beq[eq:Upert]
\beasep{1.25}{l}
i{\dot U}_{\mathrm{pert}}(t) = 
\Big(\Hb(t)+\Ht(t)\Big)\Upert(t)
\\
\ket{\psi_{\rm pert}(t)} = \Upert(t) \ket{\psi_{\rm pert}(0)}
\eea
\eeq
The nominal and perturbed Hamiltonians are assumed to be members of
known sets:
\beq[eq:HnomHpert]
\beasep{1.65}{l}
\Big\{\Hb(t),t\in[0,T]\Big\}\in{\Hbf}_{\rm nom}
\\
\Big\{\Ht(t),t\in[0,T]\Big\}\in{\Hbf}_{\rm unc}
\eea
\eeq
The nominal Hamiltonian, $\Hb(t)$, typically depends in part on the
control variables with control constraints in ${\Hbf}_{\rm
  nom}$. Since the control signal generator is also subject to
uncertainties, if these are included then the control variables will
also appear in the uncertain Hamiltonian, $\Ht(t)$.
Though $\Ht(t)$ is uncertain, the set $\Hcalunc$ has specific known
characteristics, \eg, parameters or dynamics that are ``unknown but
bounded.''

\subsection{Interaction Hamiltonian}

An alternate representation of the perturbed unitary system
\eqref{eq:Upert} which exposes the robustness issues is the {\em
  interaction unitary}
\beq[eq:Uint]
R(t) = \Unom(t)^\dag \Upert(t)
\eeq
whose evolution is given by
\beq[eq:RGint]
\beasep{1.5}{rl}
i\dot R(t) &= G(t)R(t)
\\
G(t) &= \Unom(t)^\dag\Ht(t)\Unom(t)
\eea
\eeq
where $G(t)$ is the {\em interaction Hamiltonian}. The latter shows
how the controls, acting to produce the nominal unitary, can affect
the uncertain Hamiltonian. 
Note that, if $R(0)=I$ and $G(t)$ is sufficiently small (in a suitable sense), then $R(t)\approx I$ for all $t$, i.e., $\Unom(t)$ is close to $\Upert(t)$ at all times.
Thus, the deviation of the interaction unitary $R(t)$ from identity quantifies the error and will, therefore, be used to formulate robustness in the following.

\subsection{Open bipartite system}
\label{sec:SBsys}

In general the perturbed system is assumed to be in a \emph{bipartite}
Hilbert space $\Hcal=\Hcal_S\otimes\Hcal_B$ with system space $S$ of
dimension $\ns$ and bath-space $B$ of dimension $\nb$. The resulting
Hilbert space has dimension $n=\ns\nb$. This means that any state can
be described as a linear combination of the tensor product of an
$\ns$-system basis set and an $\nb$-bath basis set.

The term ``bath' is used to denote an inaccessible environment, that
is, the bath system is very difficult to measure. Conversely, the
$S$-system is accessible by measurement.  The system-bath ($S-B$)
combination is referred to as an {\em open system}, and without the
bath, a {\em closed-system}. Though the $SB$ system is unitary, and
hence periodic, with a large enough bath the time to return is so
large as to appear irreversible. Thus the bath is a principal source
of \emph{decoherence}, \ie, from the viewpoint of the accessible
$S$-system, information is lost, and in effect, the dynamics change
from quantum to classical.

A general form of the complete Hamiltonian is,
\beq[eq:Hfull]
H(t) = H_S(t)\otimes I_B + I_S\otimes H_B(t) + H_{SB}(t)
\eeq
where $H_S(t)$ is typically the accessible part of the system, and
which includes the controls, $H_B(t)$ is the sel-dynamics of the bath,
and $H_{\rm SB}(t0$ is the couping term between the system and the
bath. If we take the nominal (uncertainty-free) Hamiltonian as
$\Hb(t)=H_S(t)\otimes I_B$, then all the uncertainty is in $\Ht(t)=
I_S\otimes H_B(t) + H_{SB}(t)$. There could of course be uncertainties
in $H_S(t)$ as well (\eg, control noise and other imperfections), and
these would then be included in $\Ht(t)$.

\subsection{Fidelity}
\label{sec:fid}

Measures of quantum computing performance are typically evaluated at a
final time $t=T$ corresponding to completion of the intended
operation. Hence, for ease of notation in this section, unless
otherwise stated, we drop the time dependence, \eg, for unitaries set
$\Upert\equiv \Upert(T), R\equiv R(T), \Unom\equiv \Unom(T)$,
\etc, and for states, set $\psiin\equiv\ket{\psi_{\rm pert}(0)}$ and
$\psif\equiv\ket{\psi_{\rm pert}(T)}$.

Let $U_{\rm des}$ denote the final-time desired
unitary. The Uhlmann-Josza fidelity \cite{Uhlmann:76,Jozsa:94}
comparing the desired output state $\psi_{\rm des}=U_{\rm des}\psiin$
with the actual output state $\psif=U\psiin$ is
\beq[eq:UJfid]
\Fuj(\psiin) 
= |\psi_{\rm des}^\dag\psif|^2
= |\psiin^\dag U_{\rm des}^\dag U\psiin|^2
\eeq
In this case $\Fuj(\psiin)$ is identical to the probability of
obtaining the desired output state for a given input state
$\psiin$.
Mathematically, $\Fuj(\psiin)$ is equal to $1$ if and only if $\ket{\psi_{\rm des}}=\ket{\psif}$.

\subsection{Performance goals} 

To evaluate performance we will use the fidelity measure \eqref{eq:UJfid}
for both the nominal (uncertainty-free) system \eqref{eq:Unom} and for
the perturbed system \eqref{eq:Upert}.  For the perturbed system there are
two goals:
\bit
\item {\em Robust performance} -- The final-time unitary $U_{\rm
  pert}$ should be close to the final-time nominal (uncertainty-free)
  unitary,
\beq[eq:Unom_des]
\Upert \approx \Unom=U_S\otimes I_B
\eeq
%

\item {\em Nominal performance} -- Let $W$ be the $S$-channel target
  unitary for $U_S$. Then the desired final-time unitary is,
  \beq[eq:Unom_fin]
  U_{\rm des} = \phi W\otimes I_B
  \eeq
  with global phase $|\phi|=1$. Thus the desired unitary is
  uncorrelated with the bath. A more general form is $U_{\rm des} =
  \phi W\otimes \Phi_B$ for any unitary $\Phi_B$. As shown in
  \cite{KosutBR:2022,KosutRabitz:24}, adding this term does allow some
  additional design flexibiity.

  \eit

\subsection{Interaction unitary}

Expressing these goals using the final-time interaction unitary,
\beq[eq:Rfin]
R = \Unom^\dag \Upert = (U_S\otimes I_B)^\dag \Upert
\eeq
gives the fidelity \eqref{eq:UJfid},
\beq[eq:UJfid phi]
\Fuj_{\rm pert}(\psiin) =
|\psiin^\dag(W^\dag U_S\otimes I_B)R\psiin|^2
\eeq
Since by definition there is no uncertainty in the model of the
nominal system, then $R_{\rm nom}=I$ and the nominal fidelity becomes,
\beq[eq:Fnom]
\Fuj_{\rm nom}(\psiin) = |\psiin^\dag(W^\dag U_S\otimes I_B)\psiin|^2
\eeq
If the target unitary $W$ is achieved by the nominal
(uncertainty-free) system, that is, $U_S=\phi W$ with global phase
$|\phi|=1$, then from \eqref{eq:Rfin}:
\beq[eq:Fid_nomactR]
\left\{
\beasep{1.5}{ll}
\mbox{nominal fidelity}
&
\Fuj_{\rm nom}(\psiin) =|\psiin^\dag\psiin|^2=1
\\
\mbox{perturbed fidelity}
&
\Fwc(\psiin) =|\psiin^\dag R \psiin|^2
\\
\mbox{interaction unitary}
&
R = (W\otimes I_B)^\dag U
\eea
\right\}
\eeq
%

\subsection{Limit of robust performance via averaging}

Clearly deviations from the nominal fidelity are captured by a
deviation of $\red{R(T)}$ from identity, \ie, if $\red{R(T)}\approx
I_n$ then $\Fwc\approx\Fnom$. This approximation was made precise in
\cite{KosutBR:2022} using the classic \emph{method of averaging}
\cite{Hale:80} and extended in \cite{KosutRabitz:24}. Define the
robustness measure as the worst-case size of the time-averaged
interaction Hamiltonian in \eqref{eq:RGint}:
\beq[eq:Jrbst]
\Jrbst =
\ds
\max_{\Hbf_{\rm unc}}\norm{\avg{G}}, 
\quad
\avg{G} =
\frac{1}{T}\int_0^T G(t)dt
\eeq
Assume the following:
\bit
\item The uncertain perturbation
Hamiltonian is bounded in the induced two-norm (the maximum singular
value)
\beq[eq:delunc]
\normsm{\Ht(t)}_2\leq\blue\delunc
\eeq
\item Simultaneously the uncertainty-free fidelity is maximized
while the robustness measure is minimized, \ie,
\beq[eq:FnomJrbst]
\beasep{1.75}{l}
\Fnom(\psiin) = 1 \mbox{ iff } U_S(T)=\phi W,\ |\phi|=1 
\\
\displaystyle
\Jrbst = \min_{\Hbf_{\rm nom}} \max_{\Hbf_{\rm unc}}
\norm{\avg{G}}
\eea
\eeq
\eit

\begin{tcolorbox}[width=\columnwidth,colframe=blue!30,colback={blue!15},title={Robust Performance Bound},outer arc=0mm,colupper=black,colbacktitle=white,coltitle=blue!100]  
  Under the conditions in \eqref{eq:delunc}-\eqref{eq:FnomJrbst},
as shown in \cite{KosutRabitz:24}, the perturbed fidelity is bounded
below by $\Flb$, \ie,
\beq[eq:Fwc4]
\Fwc(\psiin) = |\psiin^\dag R(T)\psiin|^2 \geq \Flb
\eeq
provided that,
\beq[eq:Tdel_Flb]
T(\delunc+\Jrbst)
\leq
\sqrt{4
\ln\left(
1+\sqrt{
2\left(
1-\sqrt{\Flb}
\right)
}
\right)
}
\eeq\\[-9mm]
  \end{tcolorbox}

\begin{figure}[t]
  \centering
  \includegraphics[width=0.5\textwidth]{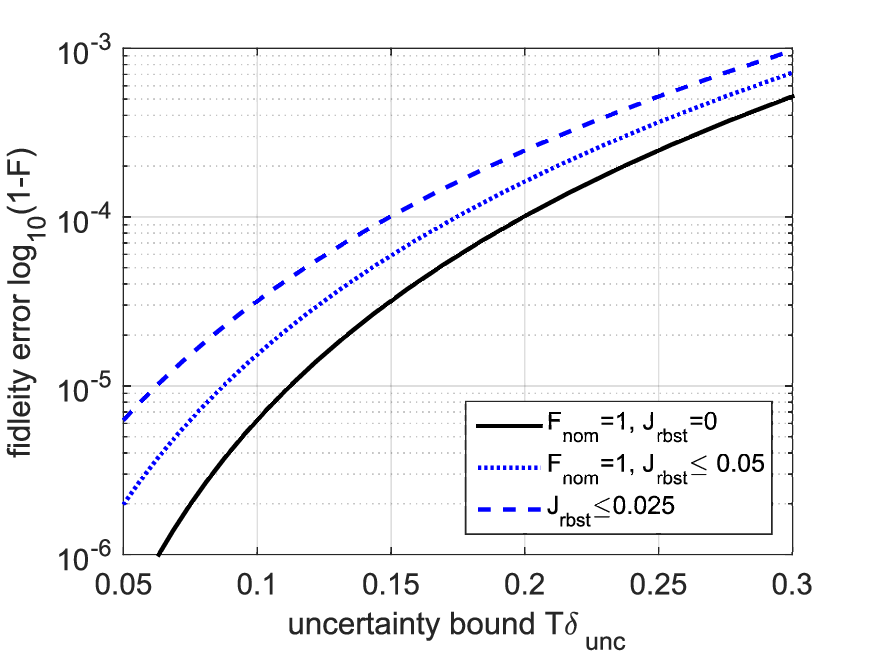}
  \caption{Performance limit curves from \eqref{eq:Tdel_Flb} shown at two
    uncertainty error scales and with a few selected robustness
    measures: solid  $\Jrbst=0$, dots $\Jrbst=0.025$, and
    dashes $\Jrbst=0.05$.}
  \label{lim}
\end{figure}

The inequality \eqref{eq:Tdel_Flb} establishes a limit of performance for
robust quantum control. This is not \emph{the} limit -- that is not
known. This performance bound does establish what is achievable under
the given conditions in \eqref{eq:delunc} and \eqref{eq:FnomJrbst}. Figure
\ref{lim} shows a plot of fidelity error $\log_{10}(1-F_{\rm pert})$
vs. uncertainty error $T\delunc$ for $\Jrbst=0$ (solid curve),
$\Jrbst=0.025$ (dotted curve), and $\Jrbst=0.05$ (dashed curve).
Some general remarks about the bound:
\bit
\item The performance bounding relationship \eqref{eq:Tdel_Flb} is based
  on three assumptions:
\bit
\item the nominal fidelity is at the maximum of $\Fnom=1$

\item the \emph{fact} that a (hopefully small) bound on $\Jrbst$ has
  been obtained, not \emph{how}.

\item the induced two-norm of the time-bandwidth product $T\Ht(t)$ is
  bounded by $\delunc$, equivalently, a bound on the product of the
  operation time and the maximum instantaneous frequency of the
  uncertain Hamiltonian.

  \eit

\item Under these conditions, the bounding relationship is:

  \bit
  \item agnostic to \emph{how} the assumptions on $\Fnom$ and $\Jrbst$
    are achieved, \eg, by control, material selection, circuit layout,
    \etc\

  \item independent of dimension.

\eit

\item Ensuring that the robustness measure is at a minimum, or at
the absolute limit of $\Jrbst=0$, clearly must account for the
specifics of the uncertainty set $\Hcalunc$. 

\eit

\subsection{Example}

\begin{figure}[h]
  \centering
  \btab{c}
  \includegraphics[width=0.5\textwidth]{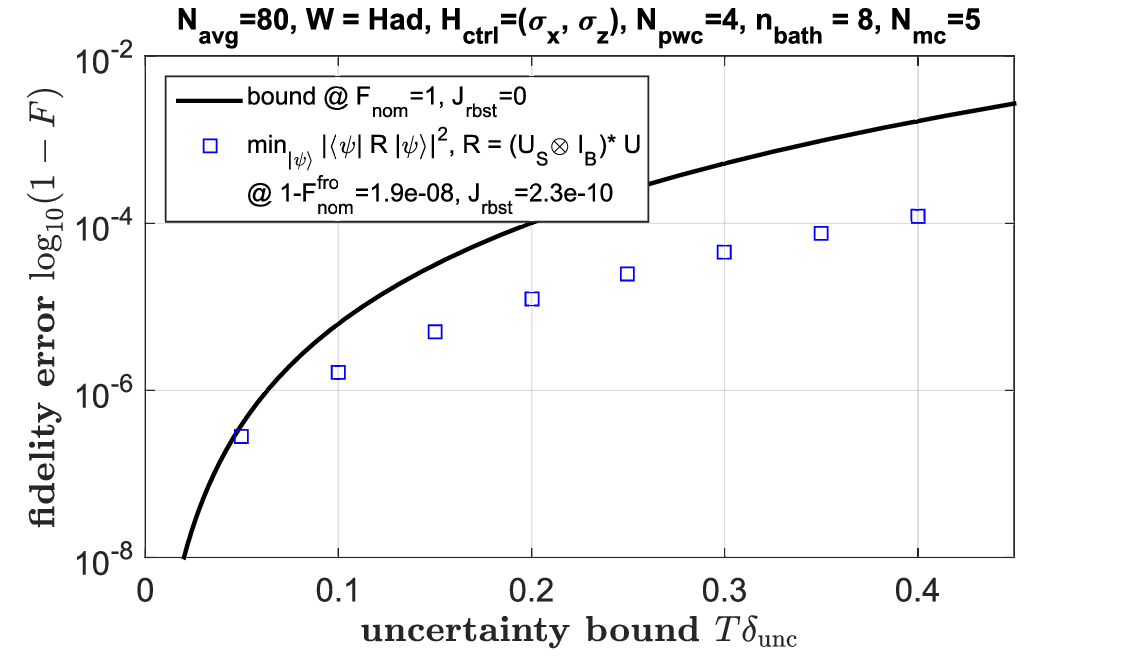}
  \\
  \includegraphics[width=0.5\textwidth]{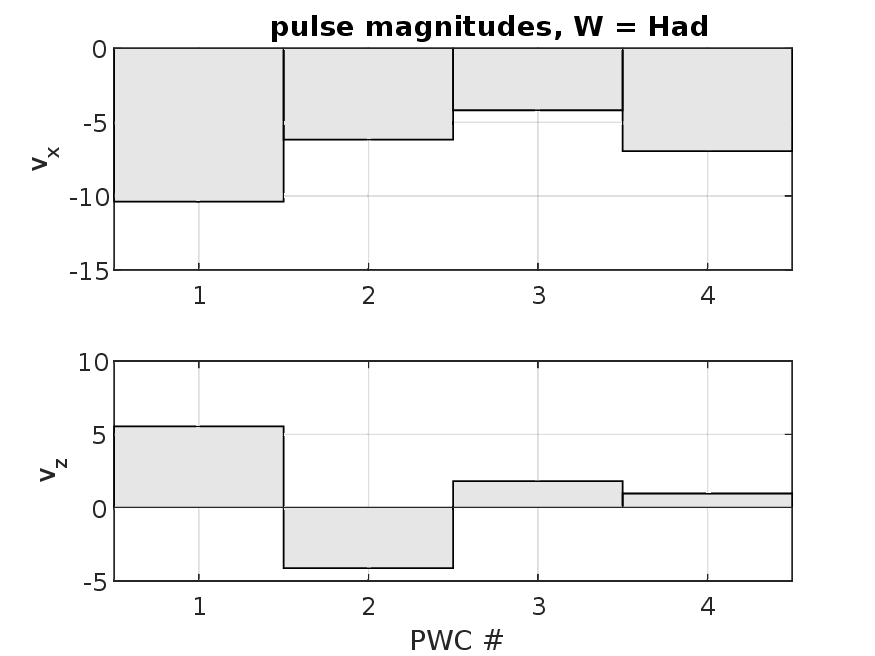}
  \etab
  \caption{{\bf Single-qubit system} \eqref{eq:Hamqu1}. Final time $T=1$,
    number of PWC pulses per control is 4 to make Hadamard. {\bf Top:}
    Solid curve is bound from \eqref{eq:Tdel_Flb}. {Blue squares} are
    actual fidelity from \eqref{eq:Fid_nomactR} from 5 random samples of the
    bath. {\bf Bottom:} control pulses.}
  \label{Ham1qu}
  \end{figure}

To illustrate the bounding relationship consider a single-qubit
system with no drift, under control in $\sig_x$ and $\sig_z$ with an
uncertain $\sig_z$ coupling $\red c_z$ to a constant uncertain bath
$\red B_z$. The system Hamiltonian is then,
\beq[eq:Hamqu1]
H(t) =
(v_x(t)\sig_x+v_z(t)\sig_z)\otimes I_B+\red
c_z\sig_z\otimes \red B_z
\eeq
Suppose all the uncertainty is due to coupling with the bath,
\beq[eq:Hamqu1 unc]
\norm{\Ht}_2 = \norm{c_z\red B_z}_2 \leq \delunc
\eeq
Set a normalized final time to $T=1$ with 4 piece-wise-constant (PWC)
control pulses each for $v_x(t)$ and $v_z(t)$. Figure \ref{Ham1qu}
shows the fidelity error and optimized control pulses to make the
Hadamard gate $W=(\sig_x+\sig_z)/\sqrt{2}$.  The solid curve is the
limit-bound and the squares show the actual worst-case fidelity from
random samples of the bath using the robust pulses. The optimization
protocol is described in the next section.

\subsection{Landscape Topology: Two-Stage Optimization}

The Averaging Theorem~\cite{Hale:80} shows that robustness can be achieved by
minimizing the size of the time-averaged Hamiltonian: the first term
in the Magnus expansion of $R(T)$. In general the \emph{trajectory} of
the nominal propagator, if properly tailored, can achieve a high
fidelity nominal response \emph{and} simultaneously minimize the
robustness measure. This latter freedom is known to arise from the
ability to roam over the null space at the top of the fidelity
landscape \toprefs.

In effect, the Averaging Theorem provides design criteria to
\emph{synthesize} a robust control.  Specifically, the final time
nominal unitary $\Unom(T)$ should be very close to the target $W$, and
the result of its evolution over $t\in[0,T]$ is that $\Jrbst\approx 0$
or as small as possible.  These two goals, which place simultaneous
demands on the \emph{nominal trajectory}, $\Unom(t),t\in[0,T]$, has been
presented in various forms in \ffrefs.

\begin{tcolorbox}[width=\columnwidth,colframe=blue!30,colback={blue!15},title={Multicriterion Optimization},outer arc=0mm,colupper=black,colbacktitle=white,coltitle=blue!100] 
  Cast as an optimization problem in the control variables:
  \beq[eq:opt_two]
  \beasep{1.25}{ll}
  \mbox{minimize}&
  \Jrbst=
  \ds
  \max_{\Hbf_{\rm unc}}\left\|
  \Gavg
  \right\|
  \\
  \mbox{subject to}&
  \Fnom \geq f_0 
  \eea
  \eeq\\[-9mm]
  \end{tcolorbox}

As described in \cite{KosutBR:2022}, one path to finding a solution to
\eqref{eq:opt_two} is to first maximize only the nominal fidelity
$\Fnom$.  When this fidelity crosses a high threshold, $f_0\approx 1$,
then switch to minimizing the robustness measure $\Jrbst$ while
keeping the fidelity above $f_0$.  Specifically, we utilize the
following two stage algorithm in the control variables $v$:
\beq[eq:twostage]
\beasep{1.25}{l}
\mbox{{\bf Stage 1} maximize {\bf only}
  $\Fnom(v)$ {\bf until} $\Fnom(v) \geq f_0$}
\\
\mbox{{\bf Stage 2} minimize $\Jrbst(v)$ {\bf while} $\Fnom(v) \geq f_0$}
\eea
\eeq
The uncertainty-free optimization in Stage 1 can be done by many
different gradient ascent methods. For Stage 2, when $\Fnom(v)\geq
f_0$, we show in \cite{KosutBR:2022} that at each iteration finding an
optimal control increment can be recast as a convex optimization
problem, \cite[App.B]{BoydV:04}.

\subsection{Concluding Remarks \& Outlook}
\label{sec:sum}

The approach to robust quantum control presented here follows
\cite{KosutBR:2022} which rests on the theoretical foundation provided
by the classic method of averaging \cite{Hale:80}.  A direct result is
a multicriterion optimization problem consisting of the
uncertainty-free fidelity competing with a generic robustness measure,
the latter being a time-domain product of a function of the controls
and the character of the uncertainty. Such a product form is
ubiquitous in the many applications of classical robust control. These
require a product in a feedback loop to be small: an uncertainty bound
and an uncertainty-free (closed-loop) map.  From the averaging theory
there naturally arises a quantitative norm measure of a time-averaged
Hamiltonian reflective of the interaction between the uncertain
perturbation and the uncertainty-free unitary evolution. This term is
equivalent to the first term in a Magnus expansion of the interaction
unitary, a term that has been used extensively as an objective for
dynamical decoupling control protocols. If this term is sufficiently
small, then at most, only second order error effects can accrue. In
the case of quantum information sciences for established realizations,
the tacit assumption is that the (possibly known or unknown)
perturbations are small.

The bound developed in \cite{KosutRabitz:24} and reproduced here in
\eqref{eq:Tdel_Flb} establishes a limit of robust control performance in
the presence of Hamiltonian set-membership uncertainty. Depending on
specifics of the uncertainty set, this limiting bound -- \emph{the}
limiting bound is not known -- relates an upper bound on fidelity
error to a bound on uncertainty magnitude (a time-bandwidth product)
as shown in Figures \ref{lim} and \ref{Ham1qu}. In general the upper
bound is agnostic to \emph{how} it is obtained, \eg, by control,
circuit layout, material selection, \emph{etc}. The information
required to make full use of the bound may require new
experiments. The goal is to expand the boundary of quantum control to
get the best performance possible.